\documentclass[preprintnumbers,amsmath,amssymb,floatfix,11pt,prd,onecolumn,superscriptaddress,nofootinbib]{revtex4}
\usepackage{diagbox}
\usepackage{latexsym}
\usepackage{epsfig}
\usepackage{epstopdf,float}
\usepackage{graphicx}
\usepackage{amssymb}
\usepackage{amsmath}
\usepackage{dcolumn}
\usepackage{bm}
\usepackage{color}
\usepackage{comment}
\usepackage[shortlabels]{enumitem}
\usepackage{subfigure}
\usepackage{multirow}
\usepackage{xcolor} 
\begin{document}
\title{\bf Visual Characteristics of a Rotating Black Hole in $4$D Einstein-Gauss-Bonnet Gravity with Thin Accretion Disk Under EHT Constraints}

\author{Muhammad Israr Aslam}
\altaffiliation{mrisraraslam@gmail.com, israr.aslam@umt.edu.pk}
\affiliation{Department of Mathematics, School of Science, University of Management and Technology, Lahore-$54770$, Pakistan.}

\author{Manahil Ali}
\altaffiliation{manomanahil600@gmail.com}
\affiliation{Department of Mathematics, University of Okara, Okara-56300 Pakistan}

\author{Sultan Hussain}
\altaffiliation{shdarwaish@imamu.edu.sa}
\affiliation{Department of Mathematics and statistics, College of Sciences, Imam Mohammad Ibn Saud Islamic University (IMSIU), Riyadh 11623 Saudi Arabia.}

\author{Abdul Malik Sultan}
\altaffiliation{ams@uo.edu.pk, maliksultan23@gmail.com}
\affiliation{Department of Mathematics, University of Okara, Okara-56300 Pakistan}

\author{Xiao-Xiong Zeng}
\altaffiliation{xxzengphysics@163.com (Corresponding
author)}\affiliation{College of Physics and Optoelectronic Engineering, Chongqing Normal University, Chongqing 401331, People’s Republic of China}

\begin{abstract}
This study investigates the visual characteristics of a rotating black hole (BH) within the fabric of $4$D Einstein-Gauss-Bonnet gravity illuminated with two illumination models, such as a celestial light sphere and a thin accretion disk. To visualize the BH shadow images, we use a recent fisheye camera model and ray-tracing method. And then, we focus on investigating the impact of the coupling parameter $\alpha$ and the spin parameter $a$ on the shadow images.  The results exhibit that the shadow radius decreases, while the shadow deviation increases with the aid of $\alpha$. However, with respect to $a$, the shadow radius is slightly increased compared to the corresponding shadow deviation. For a celestial light sphere, the increasing values of $\alpha$, lead to a decrease in the corresponding photon ring, while the space-dragging effect becomes more prominent with increasing $a$. For a thin accretion disk, we enhance its inner edge to the BH event horizon, and the particle motion is different in the regions inside and outside the innermost stable circular orbit. The result demonstrates that the shadow becomes progressively asymmetric with $a$, while the overall size of the inner shadow gradually decreases with the variations of $\alpha$. Subsequently, we also investigated the distinct features of red-shift configurations on the disk for both direct and lensed images. Additionally, we used the latest observational data from M$87^{\ast}$ and Sgr A$^{\ast}$ to impose certain parameter constraints on $\alpha$; the results depict the consistency of our considering the BH model. 
\end{abstract}
\maketitle

\section{Introduction}\label{intro}

The remarkable success of Einstein’s General Relativity (GR) in describing gravitational phenomena ranging from solar-system dynamics to gravitational wave detections has established it as the cornerstone theory of gravity \cite{1}. Several fundamental issues, such as spacetime singularities, including the primordial Big Bang singularity, and the absence of a consistent quantum description of gravity, indicate that GR may not represent the final theory of gravitation \cite{2}. These limitations have motivated sustained theoretical and observational efforts in recent decades to explore extensions of GR. In this context, modified theories of gravity provide a broader framework for addressing fundamental cosmological and high-energy phenomena beyond the standard formulation \cite{3}. Consequently, investigating gravitational models beyond GR has become essential for a deeper understanding of the underlying nature of spacetime. These considerations have motivated an extensive exploration of modified theories of gravity. Among the various extensions, Einstein–Gauss–Bonnet (EGB) gravity is particularly appealing because it incorporates higher-curvature corrections that arise in the low-energy limit of string theory \cite{4}. Although the GB term is purely topological in standard $4$-dimensional (4D) GR, a consistent 4D formulation has been proposed that produces non-trivial gravitational dynamics \cite{5}. This framework provides a promising setting for studying compact objects, especially rotating BHs, in the presence of higher-curvature effects \cite{6}.

Over the past several decades, numerous modified theories of gravity have been proposed and investigated to address theoretical and observational issues that remain difficult to explain within the framework of GR. In the context of BH physics, an important concept is the no-hair theorem, which states that within Einstein–Maxwell theory, an isolated BH is completely characterized by only three parameters: its mass, electric charge, and angular momentum, while additional non-trivial matter configurations cannot persist outside the horizon \cite{7,8}. Since its introduction by Wheeler, this idea has stimulated extensive research aimed at examining the validity and possible extensions of the theorem. In particular, the no-hair conjecture has been explored in several alternative theories of gravity, including Brans–Dicke theory, scalar–tensor models, and their broader generalization known as Horndeski gravity \cite{9a,9,10,11,12,13,14,15,16,16a,17}. Interestingly, several counterexamples to the classical no-hair theorem have been identified beyond the Einstein–Maxwell framework. In various alternative gravity models, BH solutions carrying additional fields, such as Skyrme, dilaton, and Yang–Mills–Higgs fields, have been reported. Furthermore, phenomena like spontaneous scalarization, which arise from non-minimal couplings between scalar fields and curvature invariants (for example, the Maxwell invariant, GB term, or Ricci scalar), can generate scalar-hairy BH configurations \cite{18,19,20,21,22,23,24}. These results indicate that BHs in extended gravitational theories may exhibit richer structures than those predicted by standard GR.

In astrophysical environments, the inflow of matter toward BHs frequently leads to the formation of thin accretion disks, whose structure is governed by gravitational dynamics and dissipative processes. A theoretical foundation for describing such systems was initially established by Shakura and Sunyaev using a Newtonian approach \cite{25}, and subsequently reformulated within the framework of GR by Novikov and Thorne \cite{26}. The disk description is treated as a steady configuration with a radially constant accretion rate, where the orbiting material follows nearly Keplerian trajectories. The emitted radiation is typically approximated as thermal blackbody emission arising from local equilibrium conditions in the disk. Key physical quantities, including the distribution of energy flux and radiative efficiency, which measure the conversion of the remaining mass into outgoing radiation, have been extensively analyzed in earlier studies \cite{27}.

The thin accretion disk mechanism has also proven to be a powerful tool for exploring the influence of alternative gravitational theories on observable astrophysical processes. In this regard, accretion disk properties have been examined in a wide range of modified gravity models, such as $f(R)$ gravity, scalar-tensor-vector theories, Einstein–Maxwell-dilaton models, Einstein-scalar-GB gravity, Chern-Simons gravity, and Hořava-Lifshitz gravity \cite{ligp1,ligp2,28,29,30,31,32,33,34,35,36,37}. Extensions of these analyses have been carried out in higher-dimensional scenarios, including Kaluza-Klein and brane-world constructions, as well as in the context of non-standard compact objects like wormholes, neutron stars, boson stars, fermion stars, and naked singularities \cite{38,39,40,wh1,41,42,ligp3,43,44}. Building on these developments, it becomes particularly relevant to investigate rotating BH configurations within this framework, since astrophysical BHs are expected to acquire significant angular momentum through accretion processes. In this regard, the study of disk emission and BH shadows provides a valuable means of probing the underlying spacetime structure and identifying potential deviations from GR.

In recent years, major advances in observational astrophysics have been achieved through the Event Horizon Telescope (EHT), which has successfully captured horizon-scale images of supermassive BHs, including $M87^*$ \cite{48a,48b,48c,48e,48f} and the Galactic center source $SgrA^*$ \cite{48g,48h,48i}. These observations reveal a bright emission ring surrounding a dark central region, commonly interpreted as the photon ring and the BH shadow, respectively, both of which are direct consequences of strong gravitational lensing near the event horizon. The observed shadow structure provides valuable information about the spacetime geometry and allows for precise constraints on fundamental BH parameters such as mass and spin.

The celestial light source framework has emerged as a valuable tool for investigating the optical appearance of BHs and the propagation of light in strong gravitational fields. By modeling uniform background illumination, this approach enables a clear characterization of shadow formation and spacetime geometry \cite{46,47,54,56,57}. Previous studies have demonstrated that the inclusion of rotation, magnetic fields, and observer inclination can significantly alter the observed shadow features, particularly in rotating BH spacetimes \cite{55,59}. More recently, attention has turned to the study of thin and thick accretion disks  \cite{45,israr1,50,57a}, shadows of rotating BHs \cite{49,gb1}, polarized images  \cite{51}, quasinormal modes \cite{52,jawad}, thermodynamics \cite{pop1,pop2,pop3}, accretion onto charged BH \cite{ligp4}, shadow of extreme compact charged objects \cite{ligp5} in the fabric of $4$D EGB gravity and many more found in literature. These results demonstrate that EGB gravity offers a rich framework for connecting theoretical modifications with observable astrophysical signatures. Despite these developments, a detailed analysis of the optical properties of rotating BHs in 4D EGB gravity, particularly in the presence of thin accretion disks, remains relatively limited. Since the underlying spacetime geometry directly governs the shadow and emission characteristics, they provide a valuable observational window for testing higher-curvature corrections beyond GR \cite{49}.

In order to investigate these effects, we adopt a framework similar to that employed in previous studies of BH imaging and accretion disk phenomena \cite{59,58}. Theoretically, the shadow is an inherent characteristic
of BH spacetime, it also serves as an experimental platform for testing GR and alternative gravitational theories in the context of strong gravitational fields. In present work, we consider a rotating BH solution in 4D EGB gravity and analyze the motion of photons in this spacetime by integrating the null geodesic equations. The observational appearance of the system is constructed using a backward ray-tracing technique, in which photon trajectories are traced from the observer’s screen into the BH spacetime, allowing for an efficient determination of the shadow boundary and emission profile. Furthermore, we consider astrophysically relevant illumination scenarios, including both thin accretion disk emission and distant light sources, and examine the influence of key parameters such as the BH rotation and the GB coupling constant. The physical quantities of interest, including the disk emission profile and observable image characteristics, are evaluated within this numerical framework, enabling a direct comparison with existing results in GR and other modified gravity models. Unlike the analytical approach, the numerical method requires discretizing the image plane into a grid, where each grid point corresponds to a pixel in the resulting BH shadow image. Since rotating solutions provide a more realistic description of astrophysical BHs, while the inclusion of thin accretion disks enables a closer connection with observable phenomena. Therefore, studying the optical appearance of these systems offers a meaningful way to probe the effects of the GB coupling and to identify possible deviations from standard GR predictions.

The structure of this paper is organized as follows. In Section {\bf II}, we briefly define the background of rotating BH solution in 4D EGB along with the corresponding spacetime geometry and shadow boundary. Section {\bf III} is devoted to the analysis of shadow images produced under celestial light source illumination. In Section {\bf IV}, we present a detailed description of the thin accretion disk model along with its geometrical framework and investigate the influence of relevant parameters, including the GB coupling parameter $\alpha$ and rotation parameter $a$, on the resulting shadow images. In the same section, we also investigate the redshift distribution, lensing bands, and differences between prograde and retrograde accretion flows, and constraint the model parameters with observational data. Finally, the last section summarizes our main findings and presents concluding remarks.

\section{Review of the Rotating Black Holes in $4D$ EGB gravity and its Shadow}
The action corresponding to EGB gravity in D-dimensional spacetime can be expressed as follows \cite{49,gb1}
\begin{eqnarray}\label{1}
 \mathcal{S}_{EGB}=  \int d^D x (L_{EH}+ \alpha L_{GB})\sqrt{-g},
\end{eqnarray}
where
\begin{eqnarray}\nonumber
 L_{EH}=R,\,  L_{GB}=  R^{\beta \gamma \tau \chi}-R_{\beta \gamma \tau \chi}-4 R^{\beta \gamma} R_{\beta \gamma}+R^2,
\end{eqnarray}
here, $g$ denotes the determinant of the metric tensor, while $\alpha$ is the GB coupling parameter. Furthermore, $R,~R_{\beta \gamma}$ and $R_{\beta \gamma \tau \chi}$ represent the Ricci scalar, Ricci tensor, and Riemann curvature tensor associated with the spacetime geometry. In $4D$ spacetime $(D=4)$, the GB term reduces to a total derivative and, therefore, does not contribute to the Einstein field equations. Recently, Glavan and Lin proposed a novel formulation of $4D$ EGB gravity \cite{5}, in which the GB coupling constant $\alpha$ is rescaled as $\alpha/(D-4)$, followed by taking the limit $D \rightarrow 4$. Within this framework, static and spherically symmetric BH solutions were successfully obtained. By performing the variation of the action Eq. (\ref{1}) with respect to the metric tensor 
$g_{\beta \gamma}$, one obtains the corresponding gravitational field equations.
\begin{eqnarray}\label{2}
    G_{\beta \gamma} + \alpha H_{\beta \gamma}=0,
\end{eqnarray}
here, $G_{\beta \gamma}$ represents the Einstein tensor, and $H_{\beta \gamma}$ is specified by the following relation.
\begin{eqnarray}\label{3}
  H_{\beta \gamma}=2 (RR_{\beta \gamma}-2 R_{\beta \tau }R^\tau_\gamma-2 R_{\beta \tau \gamma \chi}R^{\tau \chi}-R_{\beta \tau \chi \psi} R^{\tau \chi \psi }_\gamma)-\frac{1}{2} L_{GB}g_{\beta \gamma}.
\end{eqnarray}
It is important to note that deriving rotating BH solutions in $4D$ EGB gravity directly from the vacuum field equations is highly nontrivial. At present, no exact analytical solution for rotating configurations is known within this framework. Existing results are largely based on physically motivated constructions that reproduce the essential characteristics of rotating BHs. In this context, Kumar and Ghosh employed the Newman–Janis algorithm on a static solution in $4D$ EGB gravity to obtain a stationary and axisymmetric rotating BH metric \cite{49}. Expressed in Boyer-Lindquist coordinates, the resulting spacetime is given by the following form  \cite{49,gb1}
\begin{eqnarray}\nonumber
ds^2&=&-\bigg(\frac{\Delta-a^2 \sin^2\theta}{\Sigma}\bigg)dt^2+\frac{\Sigma}{\Delta}dr^2-2 a \sin^2\theta \bigg(1-\frac{\Delta-a^2 \sin^2\theta}{\Sigma}\bigg)dt d\phi\\ \label{4}&+&\Sigma d\theta^2+\sin^2\theta\bigg[\Sigma +a^2 \sin^2\theta\bigg(2-\frac{\Delta-a^2 \sin^2\theta}{\Sigma}\bigg)\bigg]d\phi^2,
\end{eqnarray}
along with
\begin{eqnarray}\label{5}
 \Delta= r^2 +a^2+\frac{r^4}{2 \alpha}  \bigg[1-\sqrt{1+\frac{8 \alpha M}{r^3}}\bigg],\, \Sigma=r^2+a^2\cos^2\theta,
\end{eqnarray}
where, $M$ denotes the mass of the BH. In the limit $\alpha$, the metric reduces to the rotating solution of GR, namely the Kerr spacetime. For 
$a=0$, it recovers the static, spherically symmetric BH solution in $4D$ EGB gravity. Furthermore, when both $\alpha\rightarrow0$ and $a\rightarrow0$, the metric simplifies to the Schwarzschild solution.

\begin{figure}
\centering
\subfigure[\tiny][~$\alpha=0.001$]{\label{a1}\includegraphics[width=5.4cm,height=5cm]{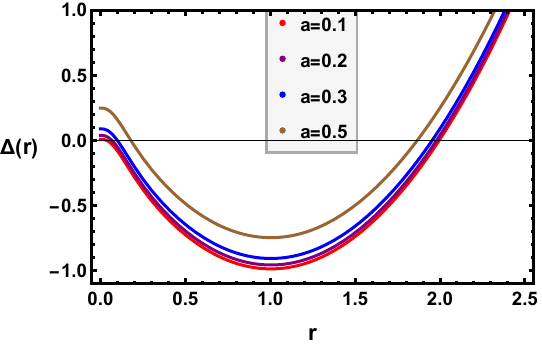}}
\subfigure[\tiny][~$\alpha=0.1$]{\label{b1}\includegraphics[width=5.4cm,height=5cm]{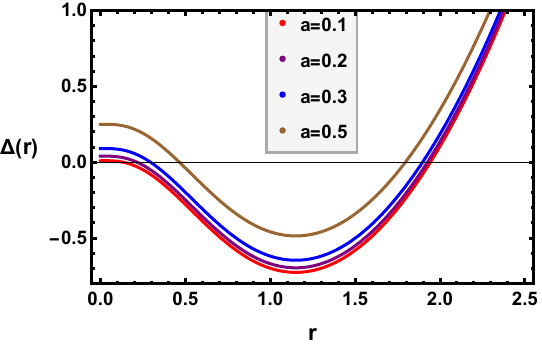}}
\subfigure[\tiny][~$\alpha=0.2$]{\label{c1}\includegraphics[width=5.4cm,height=5cm]{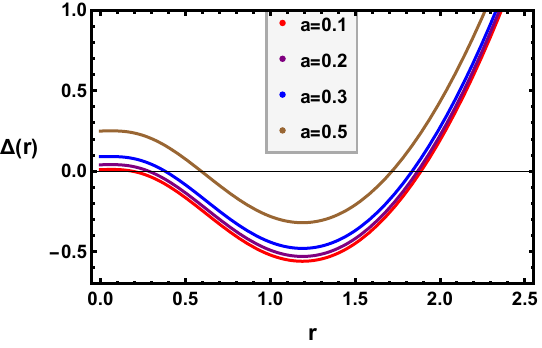}}
\subfigure[\tiny][~$\alpha=0.3$]{\label{a1}\includegraphics[width=5.4cm,height=5cm]{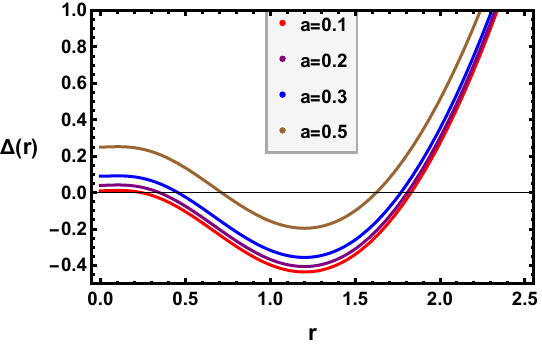}}
\subfigure[\tiny][~$\alpha=0.4$]{\label{b1}\includegraphics[width=5.4cm,height=5cm]{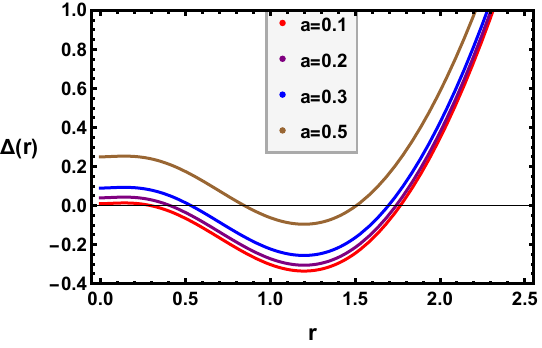}}
\subfigure[\tiny][~$\alpha=0.5$]{\label{c1}\includegraphics[width=5.4cm,height=5cm]{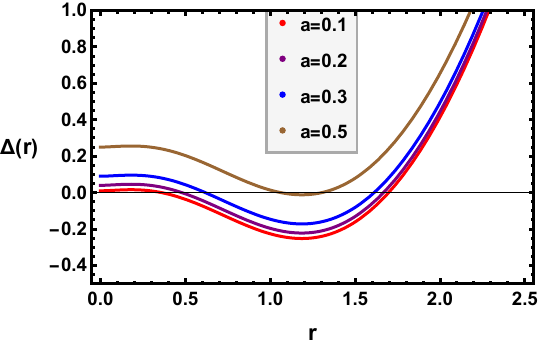}}
\subfigure[\tiny][~$\alpha=0.6$]{\label{a1}\includegraphics[width=5.4cm,height=5cm]{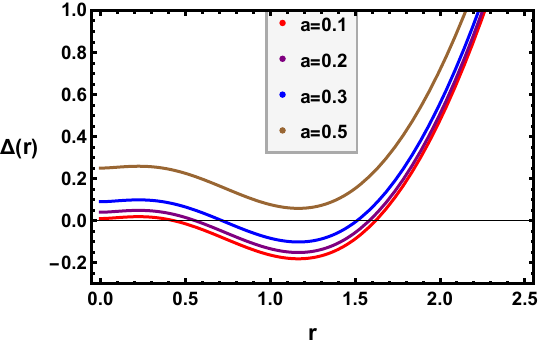}}
\subfigure[\tiny][~$\alpha=0.7$]{\label{b1}\includegraphics[width=5.4cm,height=5cm]{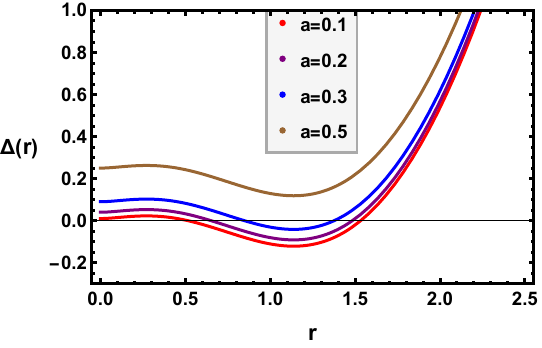}}
\subfigure[\tiny][~$\alpha=0.8$]{\label{c1}\includegraphics[width=5.4cm,height=5cm]{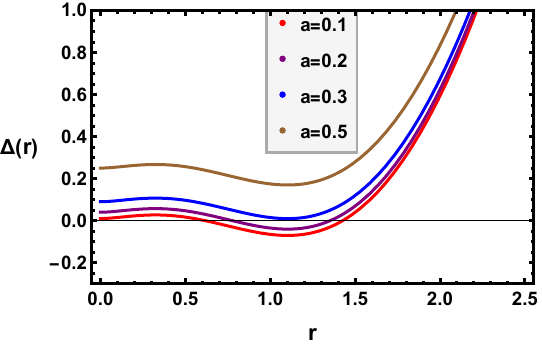}}
\caption{The physical behavior of horizons with varying BH parameters $a$ and $\alpha$. Across all panels, the red, purple, blue, and brown curves represent the cases $a=0.1,~0.2,~0.3$ and $0.5$ respectively.}\label{prd1}
\end{figure}

For a fixed value of $\alpha$, the rotation parameter 
$a$ must remain below a certain bound for the BH horizon to exist. In Fig. \textbf{\ref{prd1}}, we illustrate the behavior of $\Delta(r)$ as a function of the radial coordinate $r$ for different values of $a$,parameter $\alpha$. The results indicate the presence of two distinct horizons, namely the inner horizon (Cauchy) $r_{-}$ and the outer horizon  (event) $r_{+}$ satisfying $r_{-}<r_{+}$. As the rotation parameter approaches a critical value, the two horizons converge, resulting in an extremal BH configuration. Beyond this critical rotation, no horizon exists, indicating the absence of a BH solution. The figure shows that increasing the rotation parameter significantly affects the structure of the horizon, resulting in a decrease in the radius of the event horizon. In the limit of vanishing GB coupling, the solution approaches the Kerr BH of GR, which typically exhibits a comparatively larger horizon radius. Although the GB coupling is generally considered positive due to its connection with string tension, BH solutions have also been shown to exist for certain negative values of $\alpha$, where the singular behavior remains concealed within the horizon \cite{58ab}. In contrast to the positive case $\alpha$, condition $\Delta(r)=0$ yields only a single positive real root for the negative $\alpha$, implying the presence of a single horizon. In this work, however, we restrict our analysis to a fixed positive value of GB coupling and focus on the effects of BH rotation. It is also evident that increasing 
$\alpha$ results in an overall increase in the radius of the horizon. In the framework of 4D EGB gravity, the radius of the horizon increases with the GB coupling parameter $\alpha$.

The BH shadow is formed by photons emitted from a thin accretion disk and deflected by the strong gravitational field near the BH. Photons near the photon sphere follow unstable trajectories, where some are captured by the BH, forming the dark shadow region, while others escape and define the boundary of the shadow, known as the critical curve. Consequently, the observed shadow and disk emission provides valuable insights into the underlying spacetime geometry in EGB gravity. The geodesic dynamics of the particles is governed by the Hamilton–Jacobi equation, expressed as \cite{60}
\begin{eqnarray} \label{6}
   \frac{\partial \mathcal{A}}{\partial \lambda} = -\frac{1}{2} g^{\mu\nu} \frac{\partial \mathcal{A}}{\partial x^{\mu}} \frac{\partial \mathcal{A}}{\partial x^{\nu}}, 
\end{eqnarray}
here, $\mathcal{A}$ represents the Jacobi action of the photon and $\lambda$ denotes the affine parameter. The action $\mathcal{A}$ can be separated into the following form
\begin{eqnarray}\label{7}
   \mathcal{A} = \frac{1}{2} \sigma^2 \lambda - \mathcal{E} t + \mathcal{L} \phi + \mathcal{B}_r(r) + \mathcal{B}_\theta(\theta),
\end{eqnarray}
where we consider $\sigma=0$, which is corresponds to massless particles. The quantities $\mathcal{E}=p_t$ and $\mathcal{L}=p_\phi$ represent the conserved energy and angular momentum of the photon along the axis of rotation, respectively. The functions $\mathcal{B}_r(r)$ and $\mathcal{B}_\theta(\theta)$ depend solely on $r$ and $\theta$, respectively. Substituting the separable action into the Hamilton–Jacobi equation yields the corresponding equations of motion for photon trajectories.
\begin{eqnarray}\nonumber
 \Sigma^2 \frac{dt}{d\lambda} &=& a \big(\mathcal{L} - a \mathcal{E} \sin^2\theta \big) + \frac{r^2 + a^2}{\Delta} \big[\mathcal{E}(r^2 + a^2) - a \mathcal{L} \big], \\ \nonumber 
 \Sigma^2 \frac{dr}{d\lambda} &=& \pm \sqrt{\hat{R}(r)},\\ \nonumber
 \Sigma^2 \frac{d\theta}{d\lambda} &=& \pm \sqrt{\Theta(\theta)}, \\ \label{8}
 \Sigma^2 \frac{d\phi}{d\lambda} &=& \big(\mathcal{L} \csc^2\theta - a \mathcal{E} \big) + \frac{a}{\Delta} [\mathcal{E}(r^2 + a^2) - a \mathcal{L}],
\end{eqnarray}
along the way
\begin{eqnarray}\nonumber
   \hat{R}(r) &=& [\mathcal{E}(r^2 + a^2) - a \mathcal{L}]^2 - \Delta [J + (\mathcal{L} - a \mathcal{E})^2 ], \\ \label{9}
   \Theta(\theta) &=& J+ a^2 \mathcal{E}^2 - \mathcal{L}^2 \csc^2\theta \cos^2\theta.
\end{eqnarray}
Here, $J$ denotes the Carter constant. These equations can be used to describe the motion of photons in the vicinity of the BH. Photons moving in circular orbits form a photon sphere characterized by a radius $r_{ps}$. The condition for such circular motion requires the radial coordinate to remain constant, i.e., $\dot{r}=0$ and $\ddot{r}=0$, where the dot represents differentiation with respect to the affine parameter $\lambda$. These conditions are equivalent to $\hat{R}(r_{ps})=0$ and $\frac{d\hat{R}}{dr}|_{r=r_{ps}}=0$. The conserved quantities $\mathcal{E},~ \mathcal{L}$ and $J$ are then related to the corresponding impact parameters near the BH.
\begin{eqnarray}\label{10}
\Xi = \frac{\mathcal{L}}{\mathcal{E}}, \hspace{1cm} \zeta  = \frac{J}{\mathcal{E}^2}.  
\end{eqnarray}
The impact parameters can be determined using Eq. (\ref{10}).
\begin{eqnarray}\label{11}
\Xi(r_{ps})&=& \frac{(a^2+r_{ps}^2) \Delta' (r_{ps})-4 r_{ps} \Delta(r_{ps})}{a \Delta'(r_{ps})},   \\ \label{12} 
\zeta(r_{ps}) &=& \frac{r_{ps}^2 (-16\Delta(r_{ps})^2 - r_{ps}^2 \Delta'(r_{ps})^2 + 8\Delta(r_{ps})(2a^2 + r_{ps}\Delta'(r_{ps})))}{a^2 \Delta'(r_{ps})^2}.
\end{eqnarray}
In this notation, the prime $'$ represents a derivative with respect to $r$. The photon region is determined by the condition $\eta(r)=0$, whose solutions yield the prograde and retrograde photon orbit radii, defining the range of unstable circular photon trajectories. This region is further constrained by $\Theta(\theta)\geq 0$ for spherical photon motion. For a distant observer, modeled as a zero-angular-momentum observer (ZAMO), the BH shadow can be constructed in the image plane using a fisheye camera model and a stereographic projection technique \cite{46,58,63}. The correspondence between the four-momentum photon and the celestial coordinates $(\xi, \eta)$ then determines the observed shadow \cite{60}.
\begin{eqnarray}\label{13}
\cos\xi= \frac{p^{(1)}}{p^{0}} , \hspace{1cm}  \tan\eta= \frac{p^{(3)}}{p^{(2)}}.
\end{eqnarray}
The observer’s frame can be equipped with a standard Cartesian coordinate system 
$(x,y)$, which can be appropriately aligned with the celestial coordinates.
\begin{eqnarray}\label{14}
  x(r_{ps}) = -2 \tan(\frac{\xi}{2})\sin\eta, \hspace{0.8cm} y(r_{ps}) = -2 \tan(\frac{\xi}{2})\cos\eta.  
\end{eqnarray}

\begin{figure}
\centering
\subfigure[\tiny][~$a=0.1$]{\label{a1}\includegraphics[width=7.8cm,height=8cm]{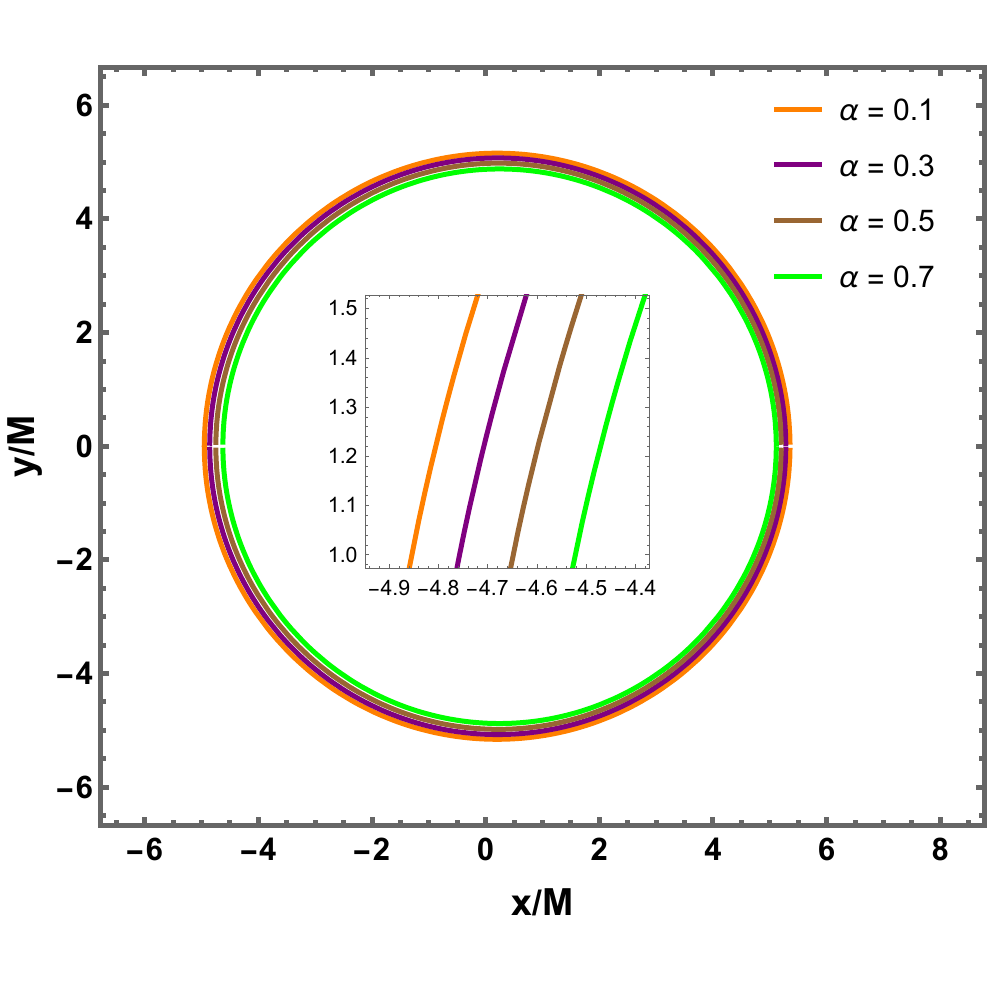}}
\subfigure[\tiny][~$a=0.35$]{\label{c1}\includegraphics[width=7.8cm,height=8cm]{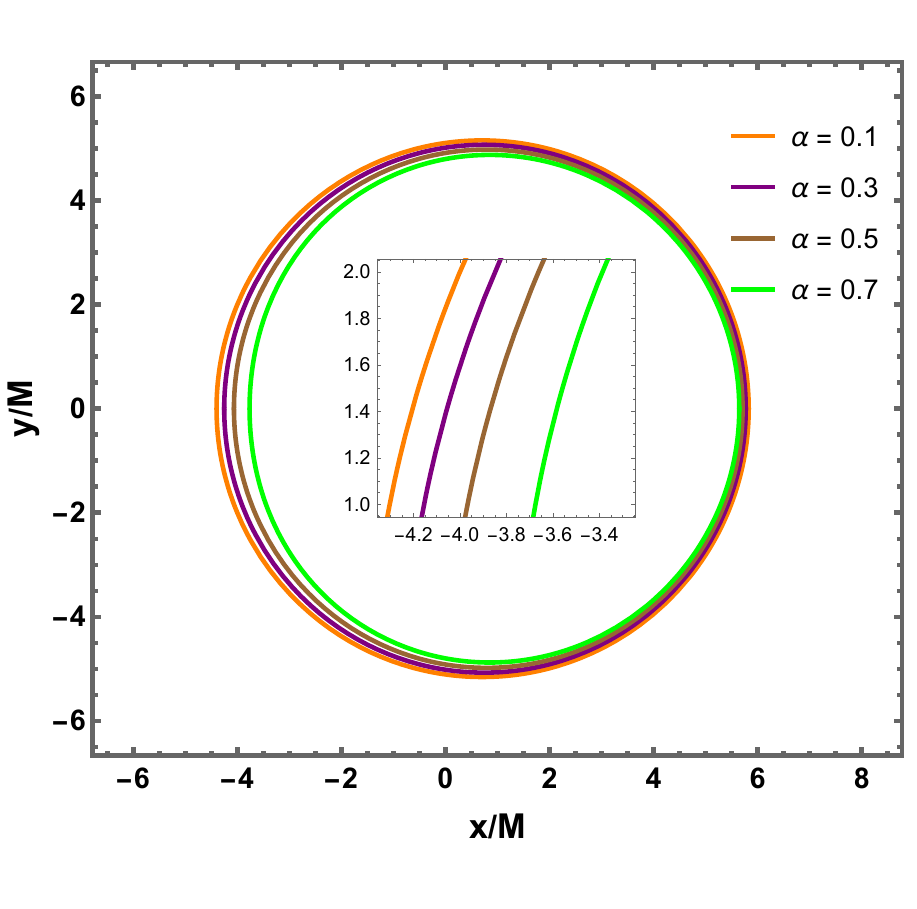}}
\caption{The shadow contours are shown in left and right panels for different values of $\alpha$ with fixed $a = 0.1$ and $a=0.35$, respectively. The observer's inclination angle is fixed at $\theta_{\text{obs}} = 80^\circ$.}\label{prd2}
\end{figure}
In Fig.~\textbf{\ref{prd2}}, we show the shadow contours for two different values of the spin parameter $a$, where the left and right panels correspond to $a=0.1$ and $0.35$, respectively. 
From the left panel ($a=0.1$), the shadow appears nearly circular, indicating that the effect of rotation is weak. The contours are symmetrically distributed, and only slight deviations from perfect circularity can be noticed. As $\alpha$ increases, the shadow radius gradually decreases, and the separation between successive curves becomes more evident in all directions. On the other hand, the right panel ($a=0.35$) shows a more distorted shadow as a result of stronger rotational effects. The contours are noticeably shifted towards the left side of the screen, demonstrating the asymmetry introduced by the spin. With increasing $\alpha$, the shadow not only decreases in size, but also exhibits a more pronounced deformation. The spacing between the curves becomes larger on the left side of the screen, whereas on the right side of the screen, the contours remain relatively closer. In general, the comparison reveals that while the GB parameter $\alpha$ mainly controls the shadow radius, the spin parameter $a$ plays a crucial role in shaping its asymmetry and deviation from circularity.

Following the approach adopted in \cite{61,62}, we describe the shadow of a rotating BH using two key observable quantities, namely the shadow radius $R_d$ and the distortion parameter $\delta_d$, which characterizes the deviation from a perfect circular shape. In this work, we follow the formalism proposed in \cite{61}, defined as follows:
\begin{eqnarray}\label{15}
  R_d= \frac{(x_t-x_r)^2+y_t^2}{2|x_t-x_r|} , \hspace{1cm} \delta_d= \frac{|x_{l'}-x_l|}{2 R_d}.
\end{eqnarray}
The size of the BH shadow can be characterized by a reference circle constructed from its top, bottom, and rightmost points, with radius $R_d$. The distortion parameter $\delta_d$ measures the deviation from circularity and is defined as the horizontal difference between the leftmost boundary of the shadow and that of the reference circle. The points $(x_t, y_t)$, $(x_b, y_b)$, $(x_r, 0)$, and $(x_l, 0)$ denote the top, bottom, right, and left positions of the shadow, respectively, while $(x_l', 0)$ corresponds to the leftmost point of the reference circle. When $x_l = x_l'$, the shadow is perfectly circular and $\delta_d = 0$; otherwise, larger values of $\delta_d$ indicate a greater deviation from circularity.

\begin{figure}[H]
\centering
\subfigure[\tiny][~$a=0.1$]{\label{a1}\includegraphics[width=7cm,height=6.5cm]{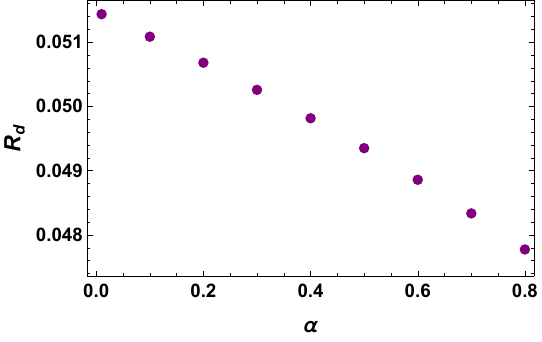}}
\subfigure[\tiny][~$a=0.1$]{\label{b1}\includegraphics[width=7cm,height=6.5cm]{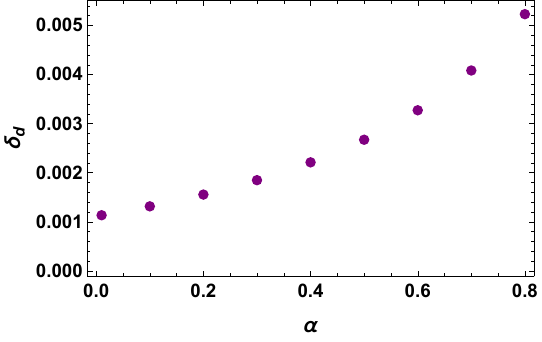}}
\subfigure[\tiny][~$\alpha=0.1$]{\label{c1}\includegraphics[width=7cm,height=6.5cm]{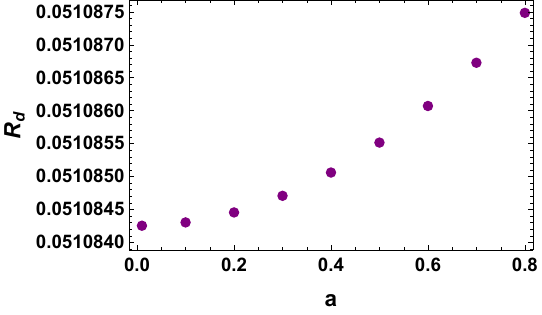}}
\subfigure[\tiny][~$\alpha=0.1$]
{\label{d1}\includegraphics[width=7cm,height=6.5cm]{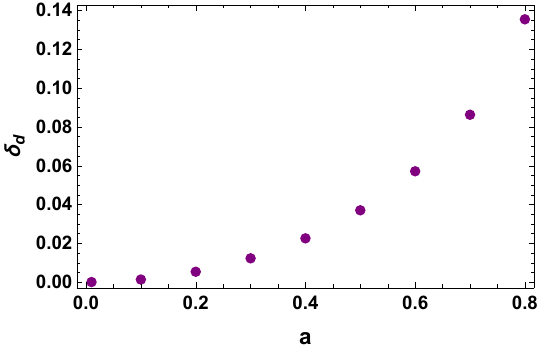}}
\caption{The behavior of $R_d$ and $\delta_d$ is illustrated for a fixed value of $a=0.1$ with varying $\alpha$ (upper panels), and for $\alpha=0.1$ with different values of $a$ (lower panels). In both cases, the observations are made for $\theta_{obs}=80^\circ,~r_{obs}=100$ and $M=1$.}\label{prd3}
\end{figure}

Figure \textbf{\ref{prd3}} illustrates the behavior of the shadow observables $R_d$ and $\delta_d$. In the upper panels, it is evident that the shadow radius 
$R_d$ gradually decreases, while the deviation parameter $\delta_d$ increases as the GB coupling $\alpha$ becomes larger. These findings are compatible with Fig.~\textbf{\ref{prd2}}, which shows that an increasing value of $\alpha$ leads to a contraction in radius of the BH shadow. In contrast, the lower panels show that both $R_d$ and $\delta_d$ increase with increasing values of the rotation parameter $a$.

\section{Shadows illuminated by the Celestial Light Source}
We employ a backward ray-tracing method to analyze the image of the BH shadow in the presence of a celestial light source. Within this framework, the BH shadow is modeled as a solid dark disk positioned at the center of the celestial sphere. Its size is considered significantly smaller than both the radius of the sphere and the separation between the observer and the origin, ensuring a realistic geometric configuration. To enhance the physical interpretation of the resulting image, the celestial sphere is partitioned into four distinct regions, each represented by a unique color (grey, lavender, teal, and  peach). These regions are defined over specific angular intervals, allowing for clear visual differentiation of photon trajectories.  In each panel, a bright circular ring is visible outside the ``D'' shaped petals, which can be interpreted as the Einstein ring. The backward ray-tracing procedure traces only the light rays that reach the observer, excluding those that are not observable. This method provides an efficient and physically meaningful way to reconstruct the apparent shadow of the BH. 

\begin{figure}
\centering
\subfigure[\tiny][~$\alpha=0.001,~a=0.001$]{\label{a1}\includegraphics[width=5cm,height=4.8cm]{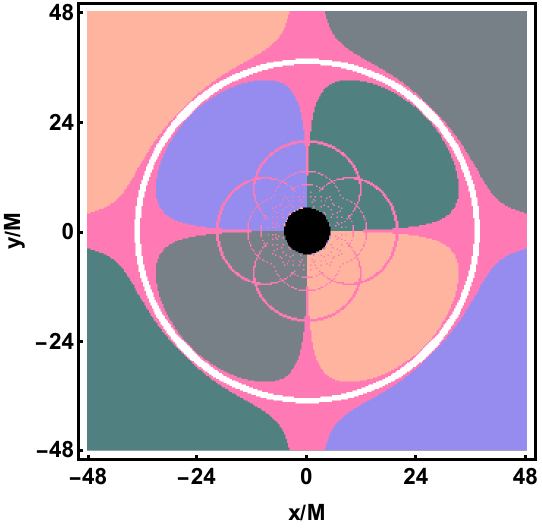}}
\subfigure[\tiny][~$\alpha=0.001,~a=0.1$]{\label{b1}\includegraphics[width=5cm,height=4.8cm]{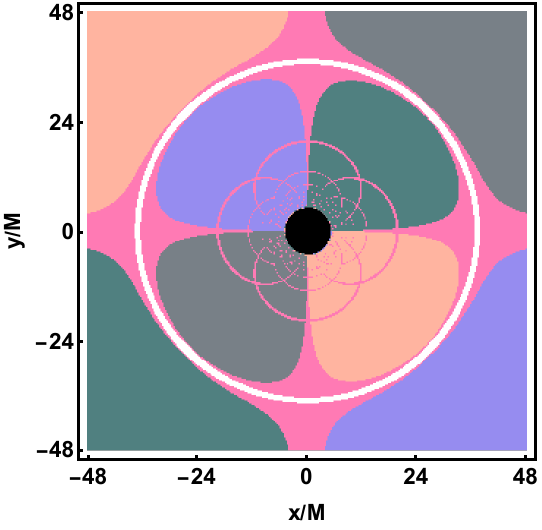}}
\subfigure[\tiny][~$\alpha=0.001,~a=0.3$]{\label{c1}\includegraphics[width=5cm,height=4.8cm]{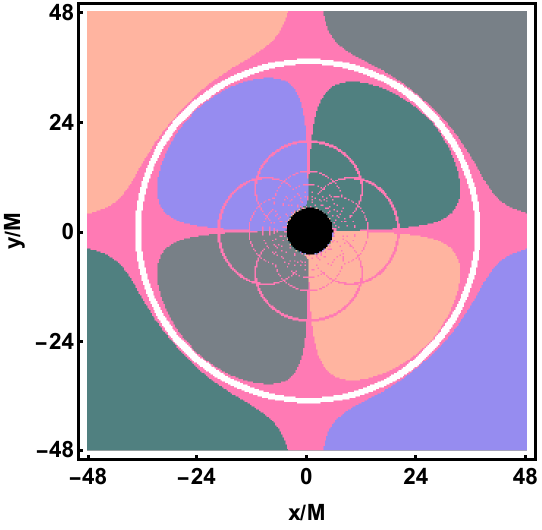}}
\subfigure[\tiny][~$\alpha=0.3,~a=0.001$]{\label{a2}\includegraphics[width=5cm,height=4.8cm]{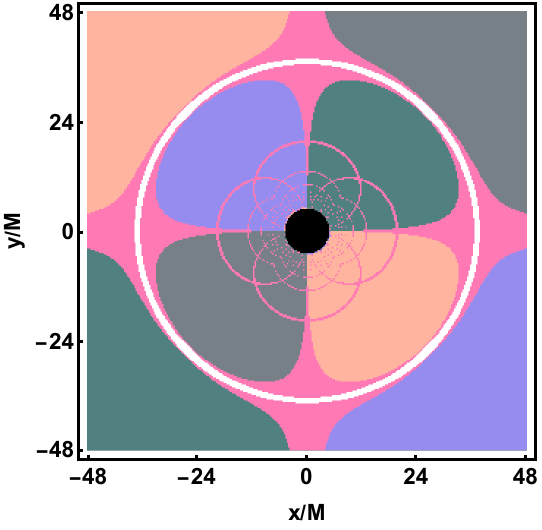}}
\subfigure[\tiny][~$\alpha=0.3,~a=0.1$]{\label{b2}\includegraphics[width=5cm,height=4.8cm]{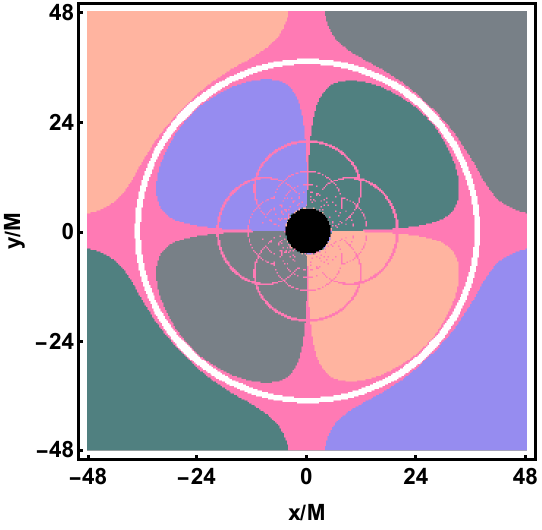}}
\subfigure[\tiny][~$\alpha=0.3,~a=0.3$]{\label{c2}\includegraphics[width=5cm,height=4.8cm]{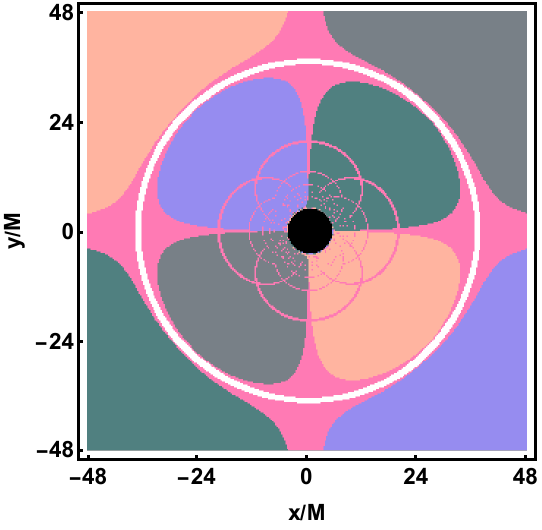}}
\subfigure[\tiny][~$\alpha=0.5,~a=0.001$]{\label{a3}\includegraphics[width=5cm,height=4.8cm]{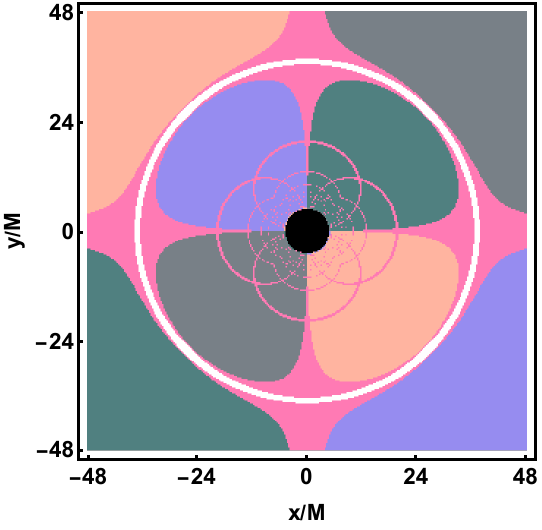}}
\subfigure[\tiny][~$\alpha=0.5,~a=0.1$]{\label{b3}\includegraphics[width=5cm,height=4.8cm]{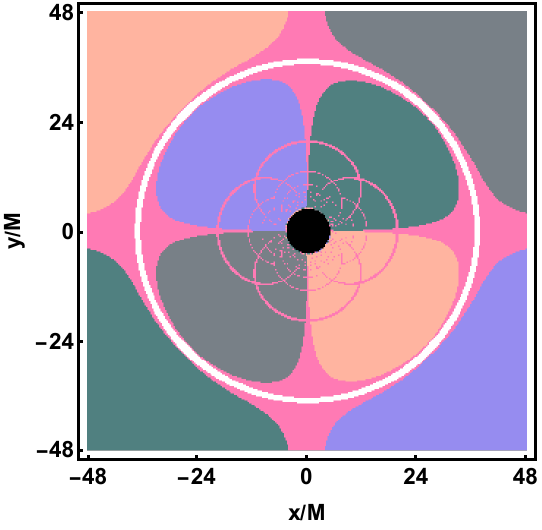}}
\subfigure[\tiny][~$\alpha=0.5,~a=0.3$]{\label{c3}\includegraphics[width=5cm,height=4.8cm]{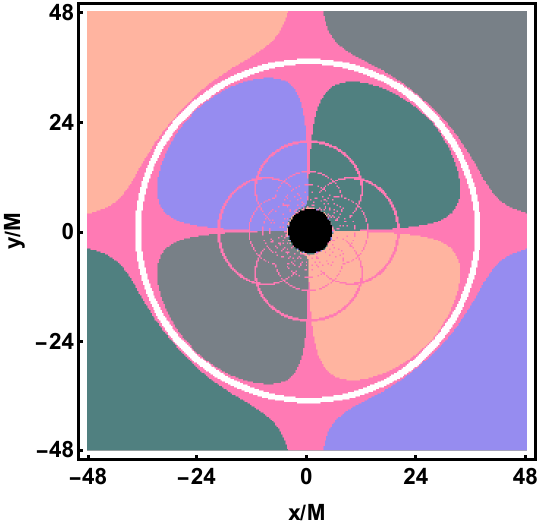}}
\subfigure[\tiny][~$\alpha=0.7,~a=0.001$]{\label{a4}\includegraphics[width=5cm,height=4.8cm]{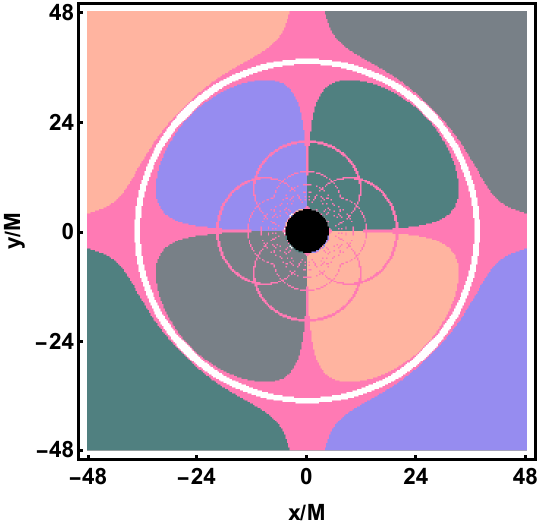}}
\subfigure[\tiny][~$\alpha=0.7,~a=0.1$]{\label{b4}\includegraphics[width=5cm,height=4.8cm]{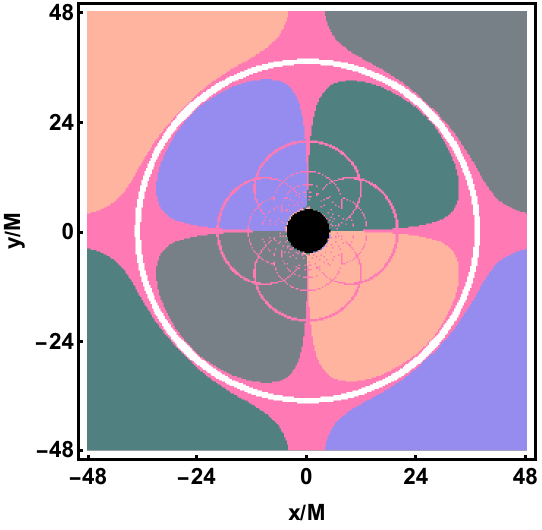}}
\subfigure[\tiny][~$\alpha=0.7,~a=0.3$]{\label{c4}\includegraphics[width=5cm,height=4.8cm]{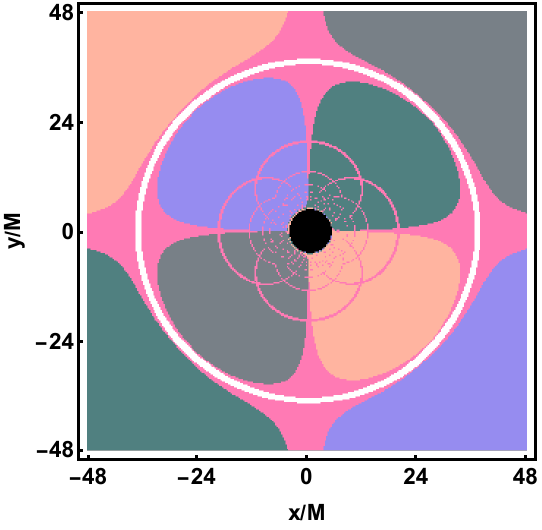}}
\caption{The BH shadow profiles corresponding to celestial light source are examined for different values of $a$ and $\alpha$, assuming an observer inclination of $\theta_{obs} = 80^\circ$. The grids of the longitude and latitude lines are marked with adjacent bright pink lines separated by $10^\circ$. The grey, lavender, teal, and peach regions represent four distinct regions of the celestial sphere. The white circular ring in each panel, which is determined by lights from the reference spot lying in the line between the BH and the observer, provides a direct interpretation of the Einstein ring. The central dark region corresponds to the BH shadow.}\label{prd4}
\end{figure}

Following the methodology outlined in \cite{58}, we further incorporate a fisheye camera projection model to simulate observational effects. Using this setup, we generate BH shadow images for various values of $a$ and $\alpha$ as illustrated in Fig.~\textbf{\ref{prd4}}. In all these images, the dark region at the center represents the BH shadow. Surrounding this shadow, a white circular structure can be seen, commonly known as the Einstein ring. 
These features clearly illustrate how a BH bends the surrounding spacetime and causes gravitational lensing, leading to the distortion of light coming from nearby sources. In Fig. \textbf{\ref{prd4}}, we analyze the shadow images of the rotating BH for different values of the spin parameter $a$ and the GB coupling constant $\alpha$. In each row, $\alpha$ is kept fixed, while $a$ increases from left to right as $a = 0.001,~ 0.1,$ and $0.3$. From the top row of Fig.~\textbf{\ref{prd4}}, it is evident that in the limiting case where both $a \rightarrow 0$ and $\alpha \rightarrow 0$, the BH shadow attains an almost perfect circular shape, which is consistent with the Schwarzschild case. As the parameter $a$ increases, the ``D'' shaped petals undergo a gradual deformation, while the radius of the surrounding bright circular ring slightly moves away from the center. Despite these changes, the size of the central dark region remains nearly unchanged across all cases.
In the second row, where $\alpha = 0.3$, a similar trend is observed. The increase in $a$ again leads to a progressive deformation of the shadow. At the same time, the radius of the bright ring surrounding the shadow becomes smaller than in the previous case, reflecting the effect of the GB parameter.
For $\alpha = 0.5$ (third row), the shadow deformation due to rotation becomes more noticeable as $a$ increases. The shadow shifts slightly and loses its symmetry more clearly than in the earlier rows. In addition, the size of the ring structure decreases further, indicating a stronger contribution from the GB coupling. Finally, in the fourth row corresponding to $\alpha = 0.7$, the combined effects of rotation and the GB parameter are more pronounced. As $a$ increases, the shadow exhibits a clearer distortion, while the surrounding ring becomes significantly larger. This suggests that higher values of $\alpha$ decrease the overall size of the observed structure, whereas the parameter $a$ mainly controls the degree of deformation. In general, the results demonstrate that the spin parameter $a$ mainly affects the shape and distortion of the shadow, while the GB parameter $\alpha$ plays a dominant role in determining the size of the surrounding ring. These features provide important insights into how rotation and modified gravity parameters influence the observable properties of BH shadows.

\section{Black Hole Shadow Formation in the Presence of Thin Disk Accretion}
The study of accretion disks surrounding BHs offers an important avenue for improving our understanding of the fundamental physical processes that govern such complex systems. In this section, we explore the observable characteristics of accretion disks within the context of $4D$ EGB gravity, where deviations from standard GR arise due to the presence of the GB coupling parameter $\alpha$. Our goal is to examine how such geometric modifications influence the observational signatures of accretion disk structures.
We consider several key aspects of the accretion model. In particular, the accreting matter is assumed to consist of electrically neutral plasma moving along equatorial timelike geodesics. Furthermore, the BH's characteristic radii play a significant role in determining both the accretion dynamics and the shadow features.
A crucial quantity in this context is the innermost stable circular orbit (ISCO), which defines the inner edge of the accretion disk. The radius of the ISCO is directly related to the efficiency of energy emission, indicating how effectively the rest-mass energy of infalling matter is converted into radiation. Specifically, the ISCO marks a transition boundary: outside this radius, matter follows stable circular orbits and contributes steadily to the emission spectrum, whereas inside this region, particles enter unstable trajectories and rapidly plunge toward the BH. The circular motion of the particles at $r_{ISCO}$ is therefore determined by the conditions required for the stability of the marginal orbit  \cite{59}.
\begin{eqnarray}\label{16}
    V_{eff}(r)=0, \hspace{1cm} \partial_r V_{eff}=0,
\end{eqnarray}
in this expression, $V_{eff}$ represents the effective potential, which is defined by
\begin{eqnarray}\label{17}
V_{eff}=(1+g^{tt}\hat{\mathcal{E}}+g^{\phi \phi}\hat{\mathcal{L}}-2g^{t\phi} \hat{\mathcal{E}}\hat{\mathcal{L}}).
\end{eqnarray}
The conserved quantities associated with the particle motion can be written as
\begin{eqnarray}\label{18}
\hat{\mathcal{E}}=-\frac{1}{\sqrt{f_{1}}}(g_{tt}+g_{t\phi}\widetilde{\omega}), \hspace{1cm}
\hat{\mathcal{L}}=\frac{1}{\sqrt{f_{1}}}(g_{t\phi}+g_{\phi\phi}\widetilde{\omega}),
\end{eqnarray}
where
$f_{1}=-g_{tt}-2g_{t\phi}\widetilde{\omega}-g_{\phi\phi}\widetilde{\omega}^{2},$ and the angular velocity is given by $
\widetilde{\omega}=\frac{d\phi}{dt}=\frac{\partial_{r}g_{t\phi}+\sqrt{(\partial_{r}g_{t\phi})^{2}-\partial_{r}g_{tt}\,\partial_{r}g_{\phi\phi}}}{\partial_{r}g_{\phi\phi}}.$
In $r=r_{ISCO}$, the corresponding conserved quantities are denoted by $\hat{\mathcal{E}}_{ISCO}$ and $\hat{\mathcal{L}}_{ISCO}$. For $r>r_{ISCO}$, the accreting matter follows stable Keplerian orbits, and its four-velocity can be expressed as
\begin{eqnarray}\label{19}
K^{\upsilon}_{out}=\frac{1}{\sqrt{f_{1}}}(1,0,0,\widetilde{\omega}).
\end{eqnarray}
Conversely, inside the ISCO, the accreting matter moves inward from the ISCO toward the event horizon along critical plunging trajectories while maintaining the conserved quantities associated with the ISCO. In this case, the components of the four-velocity are expressed as \cite{59}.
\begin{eqnarray}\nonumber
K^{t}_{plung}&=&(-g^{tt}\hat{\mathcal{E}}_{ISCO}+g^{t\phi}\hat{\mathcal{L}}_{ISCO}),\quad
K^{\phi}_{plung}=(-g^{t\phi}\hat{\mathcal{E}}_{ISCO}+g^{\phi\phi}\hat{\mathcal{L}}_{ISCO}),\\\nonumber
K^{r}_{plung}&=&-\big(-(g_{tt}K^{t}_{plung}K^{t}_{plung}+2g_{t\phi}K^{t}_{plung}
K^{\phi}_{plung}+g_{\phi\varphi}K^{\phi}_{plung}K^{\phi}_{plung}+1)(g_{rr})^{-1}\big)^{\frac{1}{2}},\\\label{20}~K^{\theta}_{plung}&=&0.
\end{eqnarray}
Light rays can intersect the accretion disk plane once $(n=1)$, twice $(n=2)$, or multiple times $(n>2)$, corresponding to direct, lensed, and higher-order images, respectively. In the present work, we restrict our analysis to the first two cases, namely the direct and lensed images. It is well established that when photons cross the accretion disk, their intensity is mainly influenced by emission and absorption processes within the disk. For the sake of simplicity, any contribution from reflection is ignored in this model. Consequently, the intensity observed on the image plane of the observer can be written as
\begin{eqnarray}\label{21}
\mathcal{I}_{obs}=\sum_{n=1}^{N_{max}}f_{n}\Psi_{n}^{3}(r_{n})\Gamma_{n}.    
\end{eqnarray}
Here, $n = 1, 2, 3, \dots, N_{\text{max}}$ denotes the number of times a photon intersects the equatorial plane, while $f_n = 1$ is a constant normalization factor. The quantity $\Psi_n = \nu_{\text{obs}} / \nu_n$ defines the redshift factor, where $ \nu_{\text{obs}}$ is the frequency measured by a distant observer, and $\nu_n$ corresponds to the photon frequency in the local rest frame comoving with the accretion disk. Furthermore, $ \Gamma_n $ is expressed as a second-order polynomial in logarithmic form and is given by
\begin{equation}\label{21}
\Gamma_{n}=\exp\big[\rho_{1}k^{2}+\rho_{2}k\big],
\end{equation}
where $ k = \log\left(\frac{r}{r_{+}}\right)$, with constants $\rho_{1} = -\frac{1}{2} $  and $\rho_{2}=-2$ \cite{nature}. Importantly, these values are consistent with the BH image captured by the EHT at a wavelength of $1.3$mm ($230$ GHz) \cite{nature}. It is important to note that the redshift factor behaves differently in the inner and outer regions of the ISCO, highlighting the significant variation in the emission characteristics of the particles in these regions. In particular, the expression for the redshift factor outside the ISCO is given by \cite{nature1}.

\begin{equation}\label{22}
\Psi^{out}_{n}=\frac{\tau_0(1-\lambda_0\frac{p_{\phi}}{p_{t}})}{\sigma_o(1+\widetilde{\omega}\frac{p_{\phi}}{p_{t}})}|_{r=r_{n}},
\quad \quad r\geq r_{ISCO},
\end{equation}
here 
$\tau_0=\sqrt{\frac{g_{\phi\phi}}{g^2_{t\phi}-g_{tt}g_{\phi\phi}}}$,~$\lambda_0=\frac{g_{t\phi}}{g_{\phi\phi}}$,~
$\sigma_o=\frac{1}{\sqrt{f_{1}}}$ and
$\bar{e}=\frac{p_{(t)}}{p_{t}}=\tau_0(1-\lambda_0\frac{p_{\phi}}{p_{t}})$ the relation between the energy detected by an observer on the screen and the energy transported along a null geodesic is characterized through a redshift factor. For an asymptotically flat spacetime, when the observer is positioned at spatial infinity, the normalized energy reduces to $\bar{e} = 1$. However, in the region where $r < r_{ISCO}$, the accreting matter follows a plunging trajectory associated with the critical orbit. In this case, the redshift factor can be expressed as \cite{nature1}
\begin{equation}\label{23}
\Psi^{plung}_{n}=-\frac{1}{K^{r}_{plung}p_{r}/p_{t}-\hat{\mathcal{E}}_{ISCO}(g^{tt}-g^{t\phi}p_{\phi}/p_{t})
+\hat{\mathcal{L}}_{ISCO}(g^{\phi\phi}p_{\phi}/p_{t}+g^{t\phi})}|_{r=r_{n}},
 \quad r< r_{ISCO}.
\end{equation}

\begin{figure}[H]\centering
    \includegraphics[width=8.5cm,height=8.5cm]{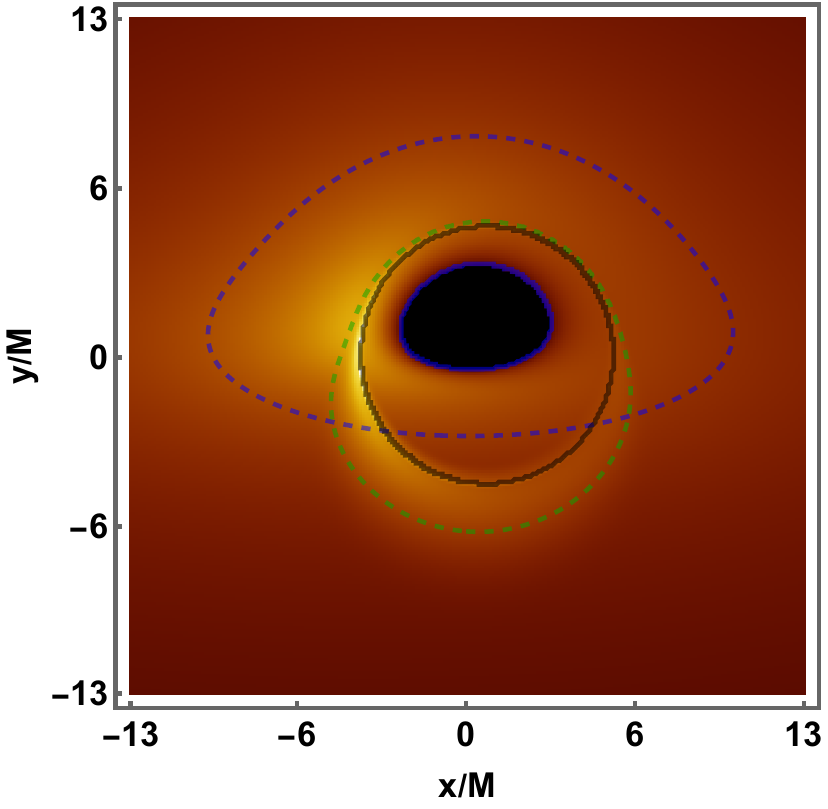}
    \caption{Physical interpretation of the four characteristic curves in the image of a rotating BH in $4D$ EGB gravity with $a=0.3$,~$\alpha=0.7$, and $\theta_{obs} = 70^\circ$ under a prograde accretion flow. The blue dashed and solid curves correspond to the direct images of the accretion disk at $r=9$ and $r=r_+$, respectively. The green dashed curve represents the lensed image of the accretion disk at $r=9$, whereas the black solid curve is the critical curve of the rotating BH.}\label{israrprd1}
\end{figure}

\begin{figure}
\centering
\subfigure[\tiny][~$\alpha=0.001,~a=0.001$]{\label{a1}\includegraphics[width=5.4cm,height=4.5cm]{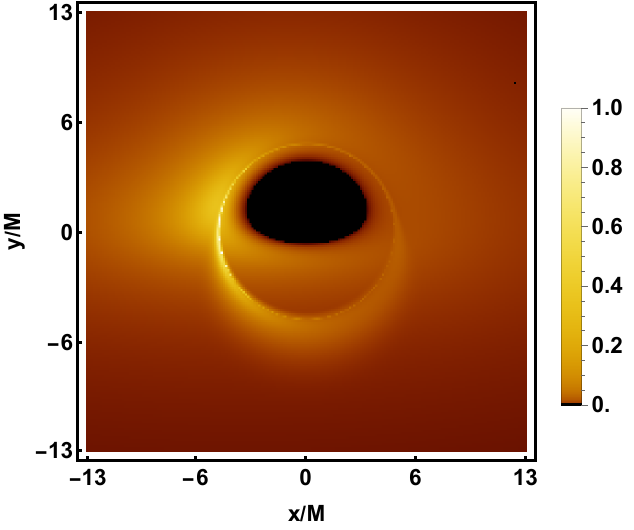}}
\subfigure[\tiny][~$\alpha=0.001,~a=0.1$]{\label{b1}\includegraphics[width=5.4cm,height=4.5cm]{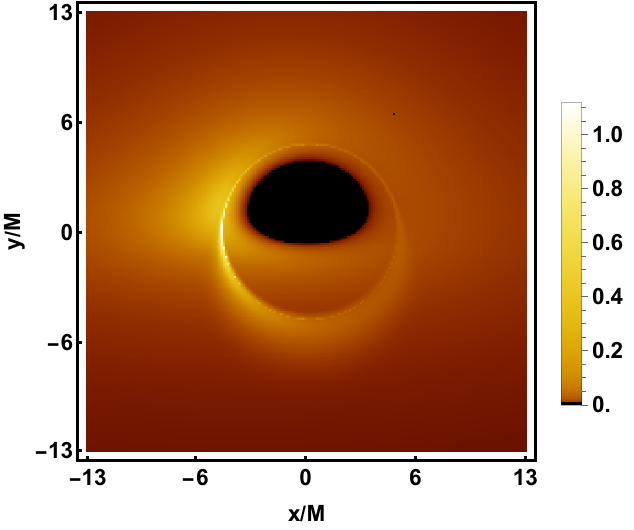}}
\subfigure[\tiny][~$\alpha=0.001,~a=0.3$]{\label{d1}\includegraphics[width=5.4cm,height=4.5cm]{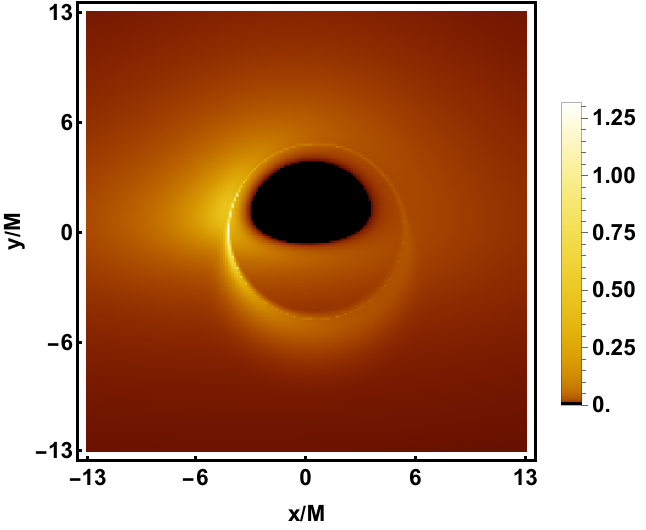}}
\subfigure[\tiny][~$\alpha=0.3,~a=0.001$]{\label{a2}\includegraphics[width=5.4cm,height=4.5cm]{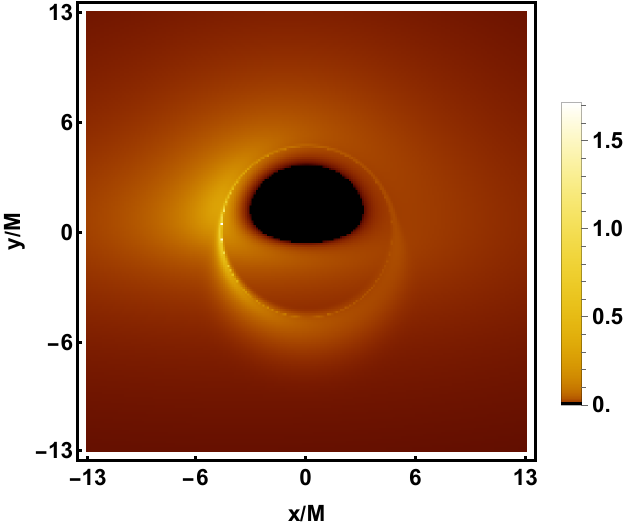}}
\subfigure[\tiny][~$\alpha=0.3,~a=0.1$]{\label{b2}\includegraphics[width=5.4cm,height=4.5cm]{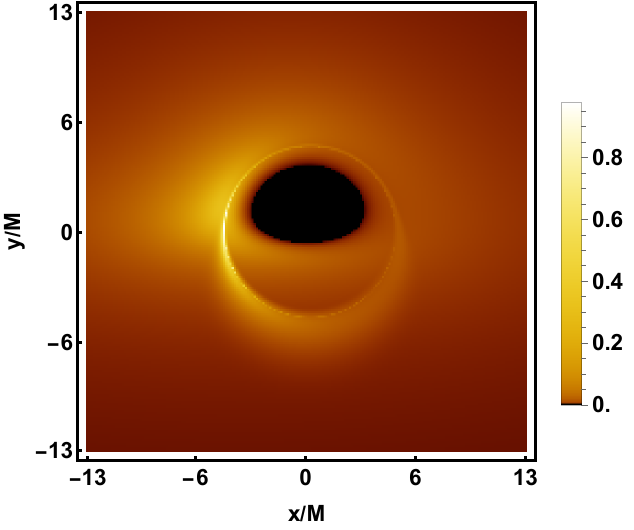}}
\subfigure[\tiny][~$\alpha=0.3,~a=0.3$]{\label{d2}\includegraphics[width=5.4cm,height=4.5cm]{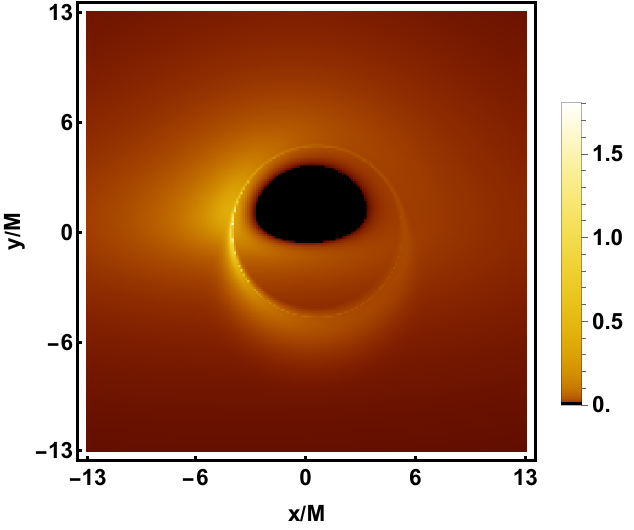}}
\subfigure[\tiny][~$\alpha=0.5,~a=0.001$]{\label{a3}\includegraphics[width=5.4cm,height=4.5cm]{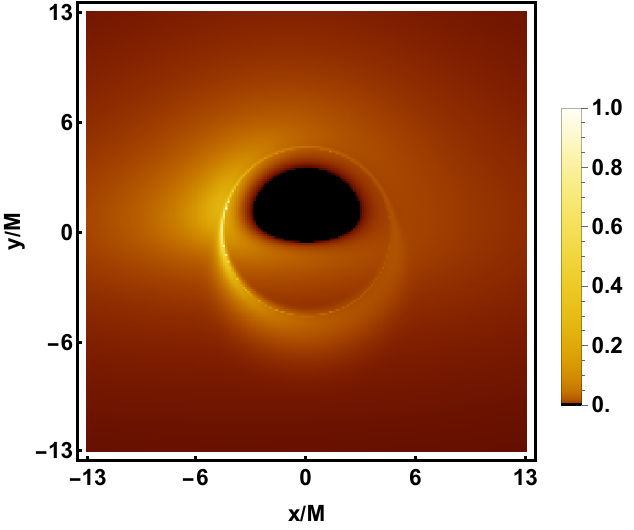}}
\subfigure[\tiny][~$\alpha=0.5,~a=0.1$]{\label{b3}\includegraphics[width=5.4cm,height=4.5cm]{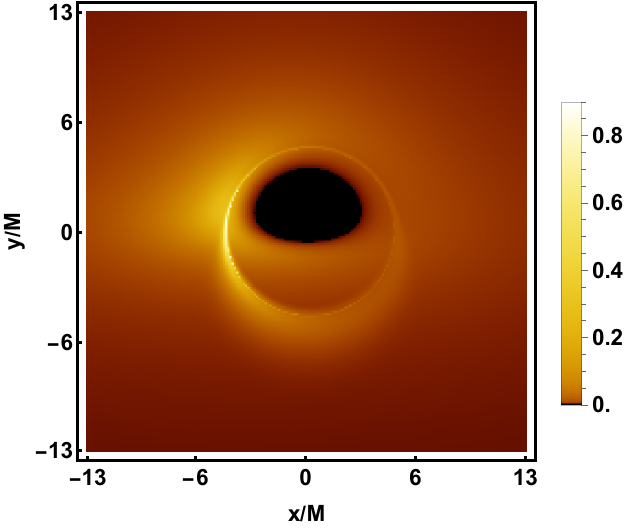}}
\subfigure[\tiny][~$\alpha=0.5,~a=0.3$]{\label{d3}\includegraphics[width=5.4cm,height=4.5cm]{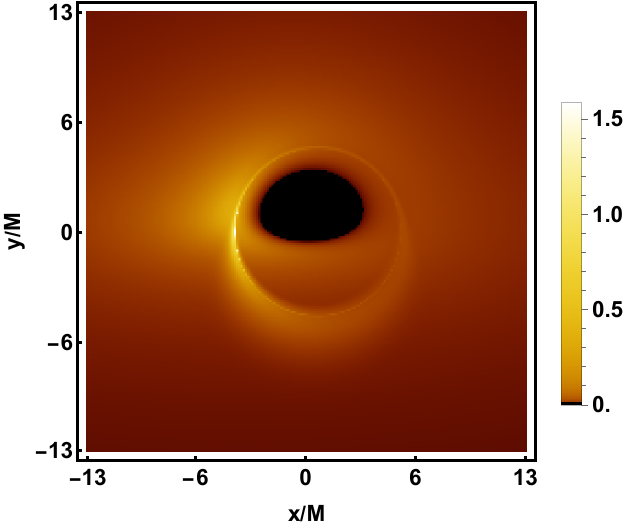}}
\subfigure[\tiny][~$\alpha=0.7,~a=0.001$]{\label{a4}\includegraphics[width=5.4cm,height=4.5cm]{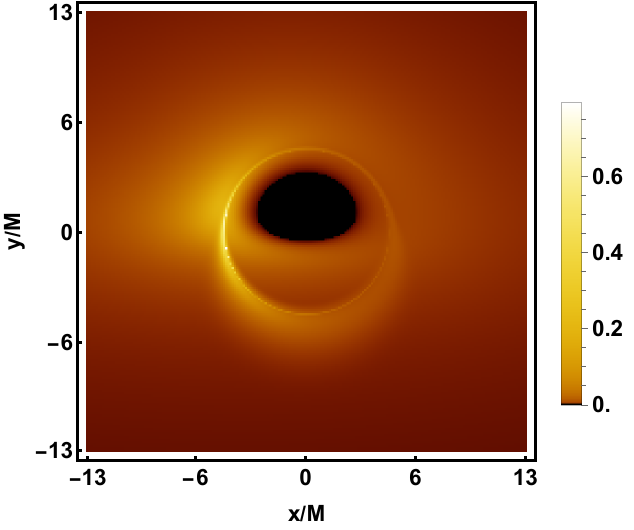}}
\subfigure[\tiny][~$\alpha=0.7,~a=0.1$]{\label{b4}\includegraphics[width=5.4cm,height=4.5cm]{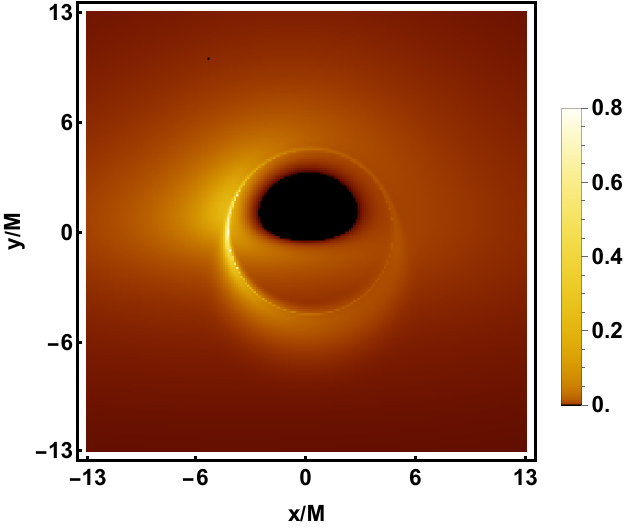}}
\subfigure[\tiny][~$\alpha=0.7,~a=0.3$]{\label{d4}\includegraphics[width=5.4cm,height=4.5cm]{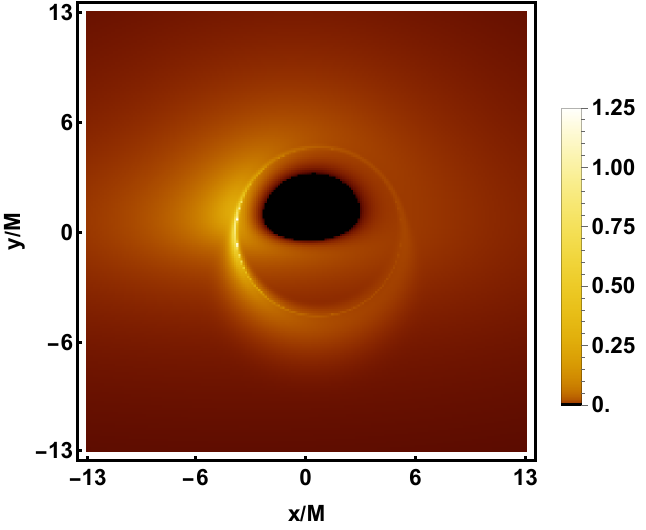}}
\caption{Optical images of rotating BHs in $4D$ EGB gravity is examined for different values of $a$ and $\alpha$, while keeping the observer inclination fixed at $\theta_{obs} = 70^\circ$ under a prograde accretion flow. The event horizon of the BH appears as a dark region, whereas the surrounding bright circular ring indicates the location of the photon ring.}\label{prd5}
\end{figure}

We present four characteristic curves for the rotating $4D$ EGB BH in Fig.~\textbf{\ref{israrprd1}}. The dashed curves represent the images of the accretion disk located at $r=9$, with the blue and green dashed curves corresponding to the direct and lensed images, respectively. The blue solid curve illustrates the image of the accretion disk at $r=r_+$, commonly known as the inner shadow of the BH \cite{nature}. The solid black curve represents the critical curve of the rotating BH, which is widely acknowledged as the photon ring in the literature. In Fig.~\textbf{\ref{prd5}}, we present the effects of the spin parameter $a$ and the GB parameter $\alpha$ for a fixed observer inclination angle $\theta_{\text{obs}} = 70^\circ$ under prograde flow. In all panels, a dark central region is clearly visible, representing the photon capture zone of the BH where light rays are unable to escape. Encircling this region, a bright closed curve appears, which corresponds to the critical curve \cite{67}. The enclosed shadow exhibits a distorted, nearly ``D'' shaped structure in each case.
To examine the influence of the GB parameter $\alpha$, we fix $\alpha$ along each row and vary the spin parameter $a$ from left to right. It is observed that as $a$ increases, the shadow becomes progressively more asymmetric. The deformation is more pronounced toward one side, and a bright crescent-like feature emerges along the boundary of the critical curve. This feature becomes increasingly prominent with larger values of $a$, reflecting the growing impact of frame-dragging.
Conversely, to study the role of spin parameter $a$, we fix $a$ along each column and vary $\alpha$ from top to bottom. With increasing $\alpha$, the overall size of the inner shadow gradually decreases, while its shape undergoes only mild deformation. Additionally, the bright emission region tends to shift slightly, indicating a redistribution of the observed intensity. The enhanced brightness on one side of the shadow is a consequence of relativistic Doppler boosting, where radiation from the approaching side of the accretion flow is blue-shifted, leading to higher observed intensity. In general, the figure highlights that $\alpha$ primarily influences the size of the shadow, while $a$ governs its asymmetry and brightness distribution. We now turn to a more detailed analysis of the BH imaging process, focusing on a precise determination of the redshift factors associated with the motion of the emitting particles. These features can be understood in terms of Doppler-related effects.
\begin{figure}
\centering
\subfigure[\tiny][~$\alpha=0.001,~a=0.001$]{\label{a1}\includegraphics[width=5.4cm,height=4.5cm]{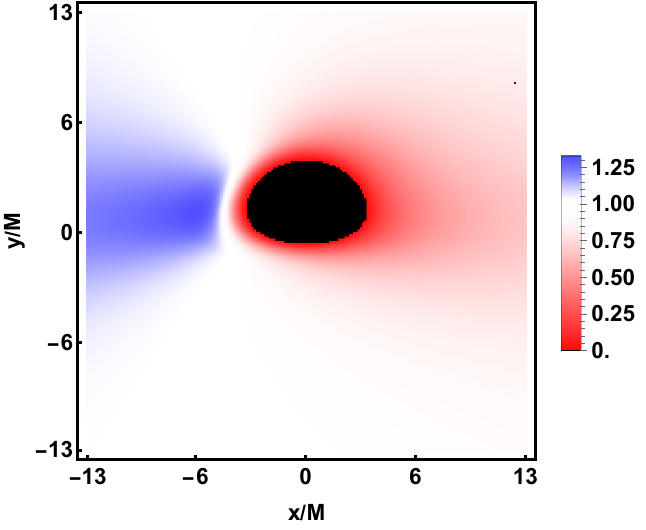}}
\subfigure[\tiny][~$\alpha=0.001,~a=0.1$]{\label{b1}\includegraphics[width=5.4cm,height=4.5cm]{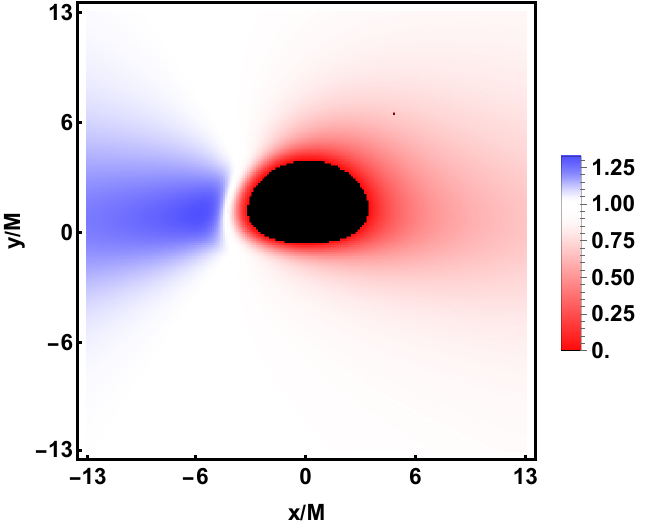}}
\subfigure[\tiny][~$\alpha=0.001,~a=0.3$]{\label{d1}\includegraphics[width=5.4cm,height=4.5cm]{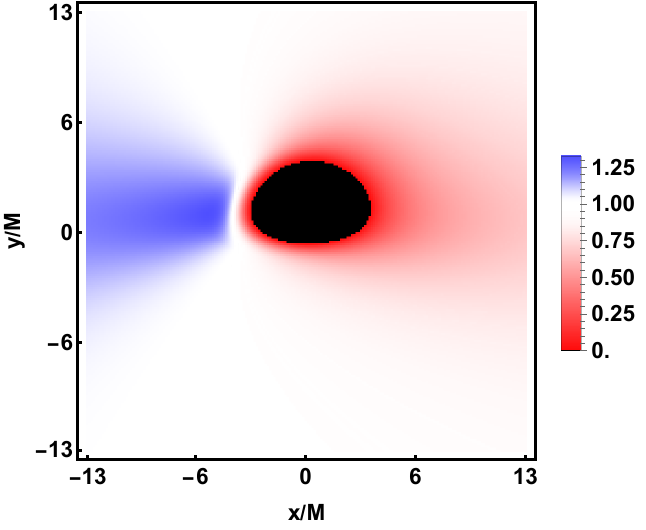}}
\subfigure[\tiny][~$\alpha=0.3,~a=0.001$]{\label{a2}\includegraphics[width=5.4cm,height=4.5cm]{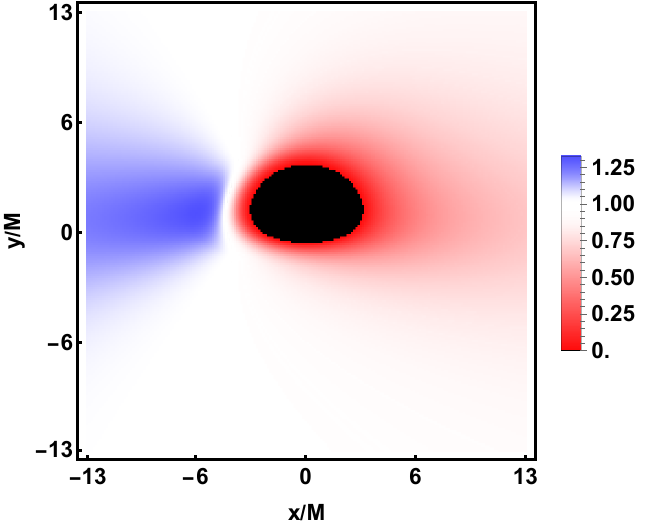}}
\subfigure[\tiny][~$\alpha=0.3,~a=0.1$]{\label{b2}\includegraphics[width=5.4cm,height=4.5cm]{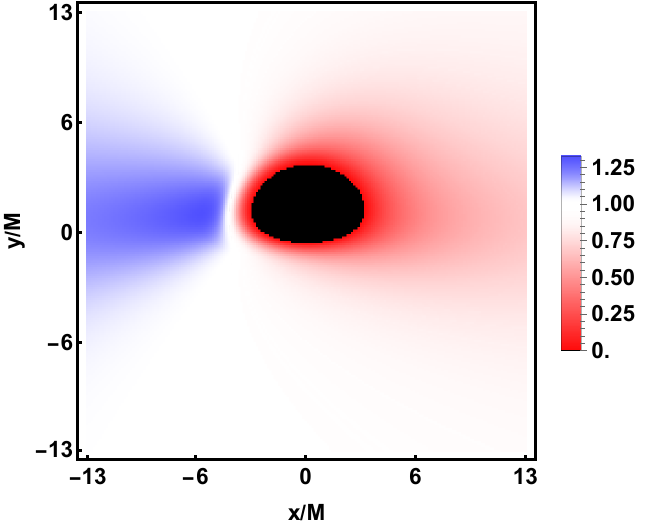}}
\subfigure[\tiny][~$\alpha=0.3,~a=0.3$]{\label{d2}\includegraphics[width=5.4cm,height=4.5cm]{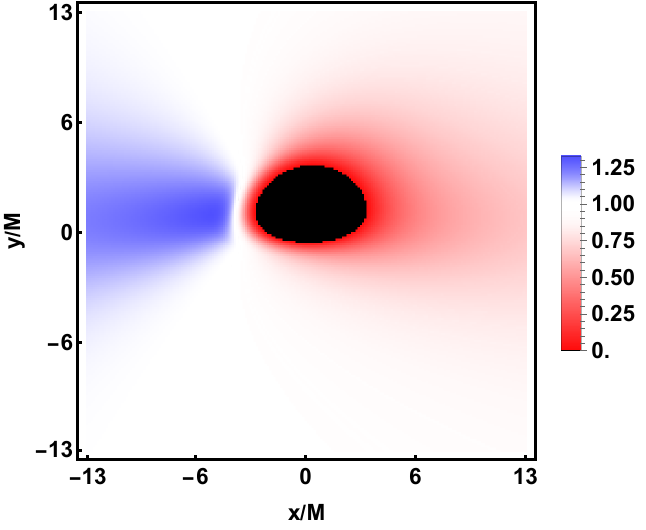}}
\subfigure[\tiny][~$\alpha=0.5,~a=0.001$]{\label{a3}\includegraphics[width=5.4cm,height=4.5cm]{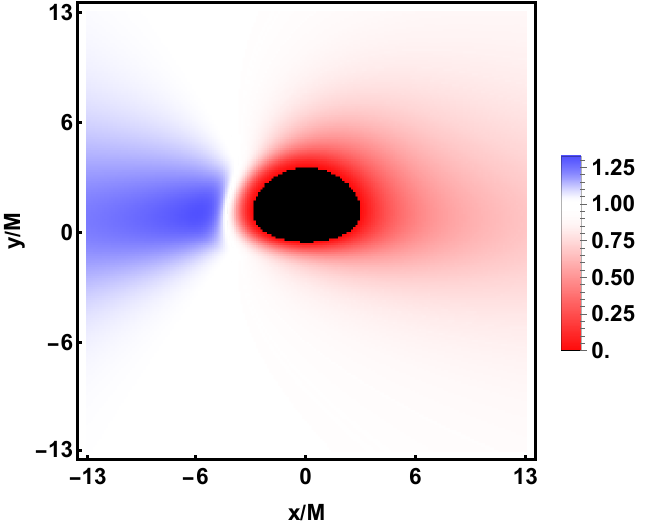}}
\subfigure[\tiny][~$\alpha=0.5,~a=0.1$]{\label{b3}\includegraphics[width=5.4cm,height=4.5cm]{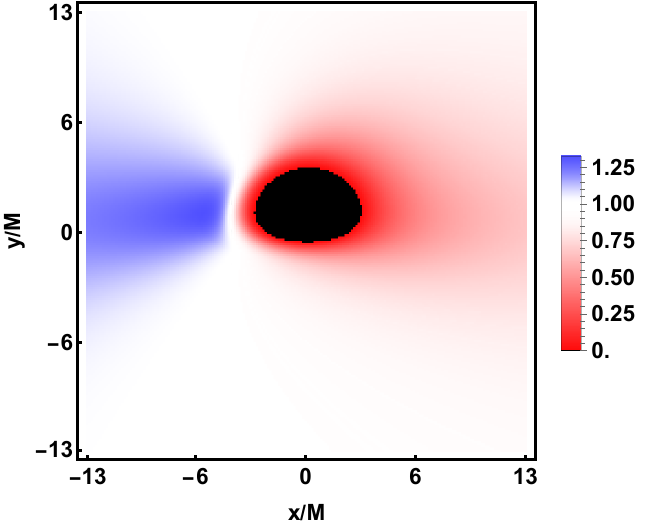}}
\subfigure[\tiny][~$\alpha=0.5,~a=0.3$]{\label{d3}\includegraphics[width=5.4cm,height=4.5cm]{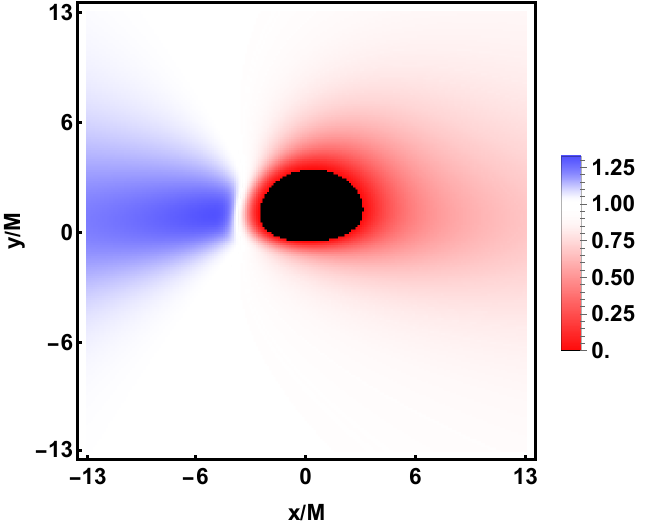}}
\subfigure[\tiny][~$\alpha=0.7,~a=0.001$]{\label{a4}\includegraphics[width=5.4cm,height=4.5cm]{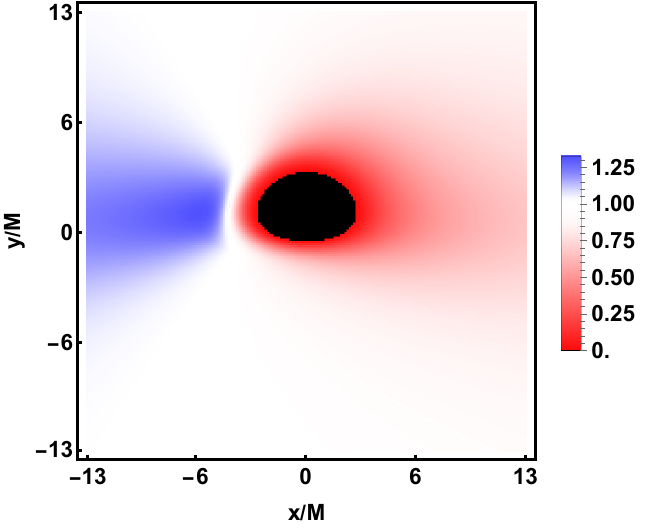}}
\subfigure[\tiny][~$\alpha=0.7,~a=0.1$]{\label{b4}\includegraphics[width=5.4cm,height=4.5cm]{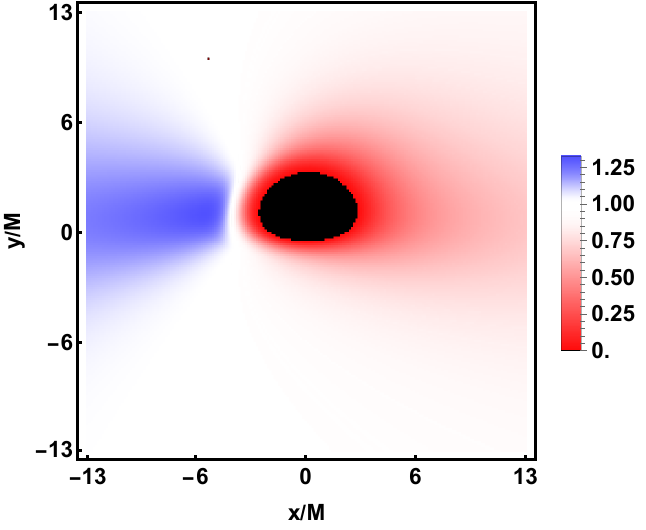}}
\subfigure[\tiny][~$\alpha=0.7,~a=0.3$]{\label{d4}\includegraphics[width=5.4cm,height=4.5cm]{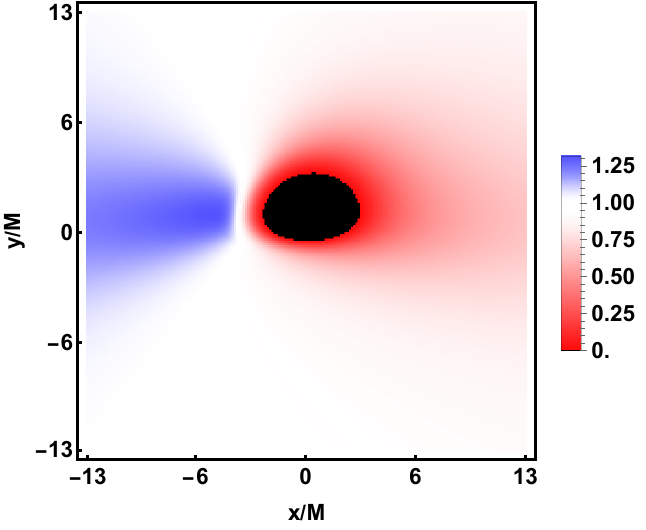}}
\caption{The redshift distribution of direct images for rotating BHs in $4D$ EGB gravity is examined for different values of $a$ and $\alpha$,  with the observer’s inclination fixed at $\theta_{obs} = 70^\circ$ under prograde motion. The red and blue regions correspond to redshifted and blueshifted, respectively.}\label{prd6}
\end{figure}

Figure~\textbf{\ref{prd6}} provides a detailed visualization of how the key parameters influence the distribution of red-shift factors in the case of prograde flow. The color contrast distinguishes the frequency shifts, where red regions represent redshifted emission and blue regions indicate blueshifted emission. In all panels, the dark central area corresponds to the inner shadow of the BH. A clear asymmetry in the frequency distribution is observed across all configurations: the blueshifted region consistently appears on the left side of the image, while the redshifted region is concentrated on the right side. This directional pattern reflects the motion of the emitting matter relative to the observer. Examining the variation along each column (from top to bottom), where the GB parameter $\alpha$ is increased, it is evident that the intensity of the red-shift gradually weakens. The corresponding redshifted region becomes less prominent, indicating a reduction in its observational strength with higher rotation. In contrast, along each row (from left to right), where the spin parameter $a$ is varied, the redshifted emission shows a slight contraction. As $a$ increases, the red shift region occupies a smaller portion of the image and its luminosity becomes marginally reduced compared to lower values. Furthermore, in all cases, the redshifted region forms a well-defined boundary surrounding the central shadow, while the blueshifted emission is located slightly farther away from this boundary. This spatial separation highlights the combined influence of gravitational and kinematic effects on the observed radiation. In general, all these results demonstrate that the GB parameter $\alpha$ strongly affects the intensity and distribution of redshift, while the spin parameter $a$ introduces more subtle modifications to its spatial extent.

\begin{figure}
\centering
\subfigure[\tiny][~$\alpha=0.001,~a=0.001$]{\label{a1}\includegraphics[width=5.4cm,height=4.5cm]{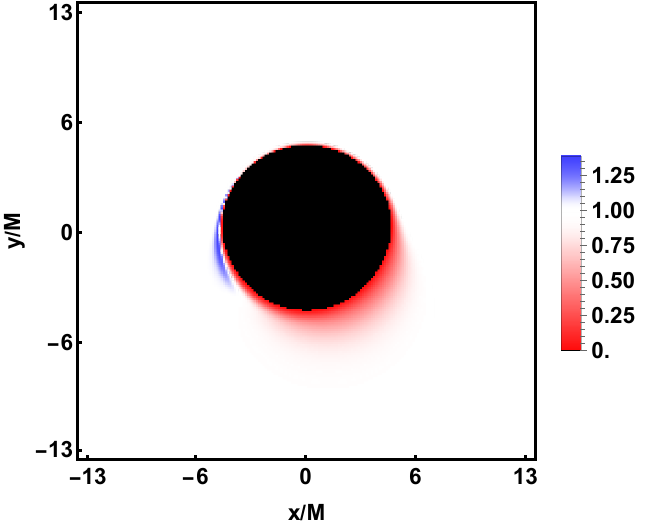}}
\subfigure[\tiny][~$\alpha=0.001,~a=0.1$]{\label{b1}\includegraphics[width=5.4cm,height=4.5cm]{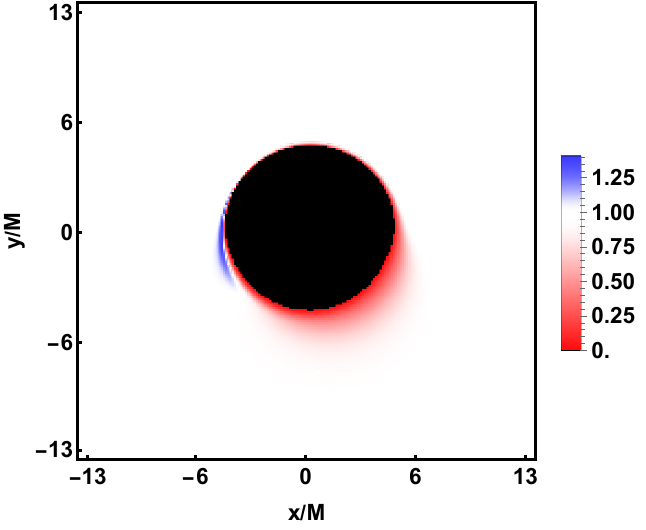}}
\subfigure[\tiny][~$\alpha=0.001,~a=0.3$]{\label{d1}\includegraphics[width=5.4cm,height=4.5cm]{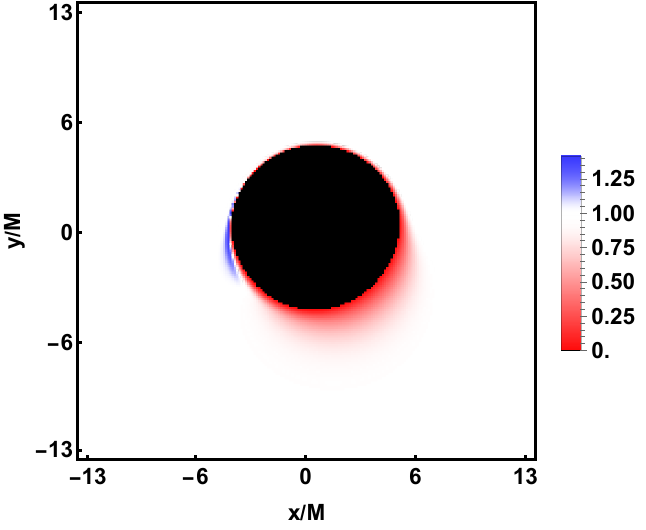}}
\subfigure[\tiny][~$\alpha=0.3,~a=0.001$]{\label{a2}\includegraphics[width=5.4cm,height=4.5cm]{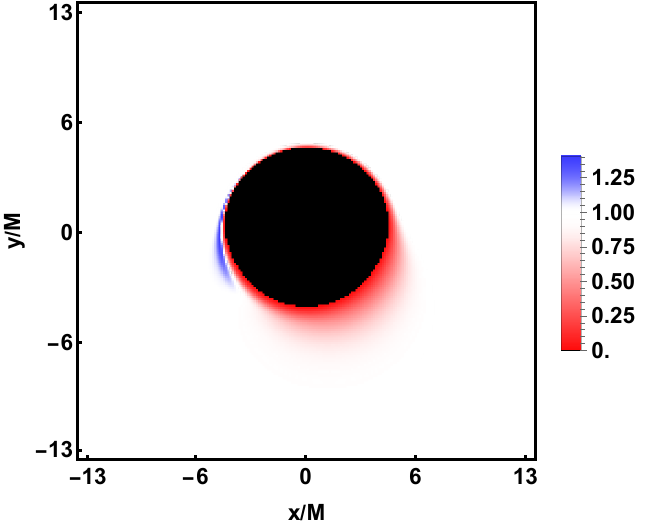}}
\subfigure[\tiny][~$\alpha=0.3,~a=0.1$]{\label{b2}\includegraphics[width=5.4cm,height=4.5cm]{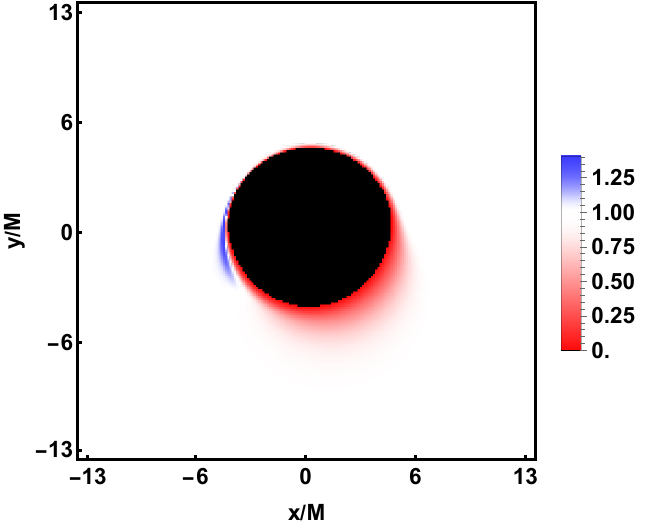}}
\subfigure[\tiny][~$\alpha=0.3,~a=0.3$]{\label{d2}\includegraphics[width=5.4cm,height=4.5cm]{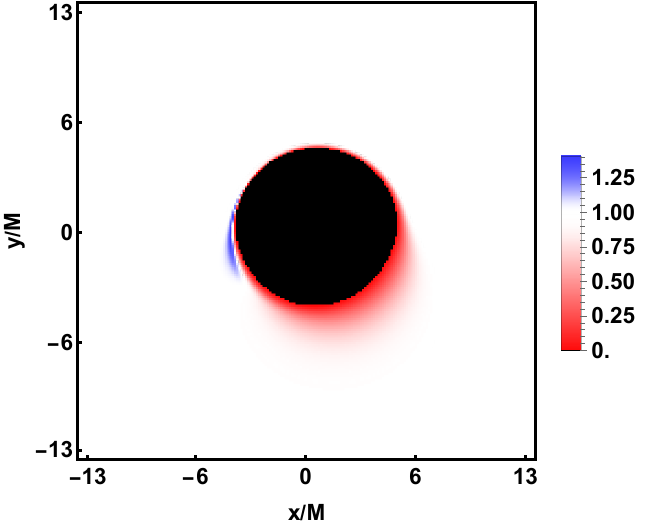}}
\subfigure[\tiny][~$\alpha=0.5,~a=0.001$]{\label{a3}\includegraphics[width=5.4cm,height=4.5cm]{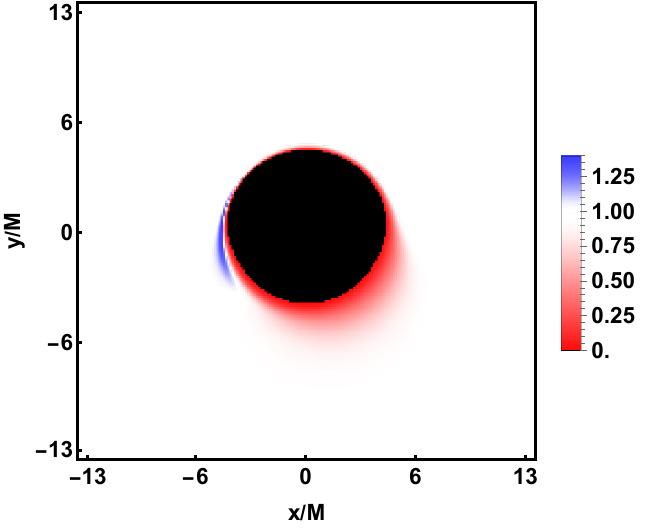}}
\subfigure[\tiny][~$\alpha=0.5,~a=0.1$]{\label{b3}\includegraphics[width=5.4cm,height=4.5cm]{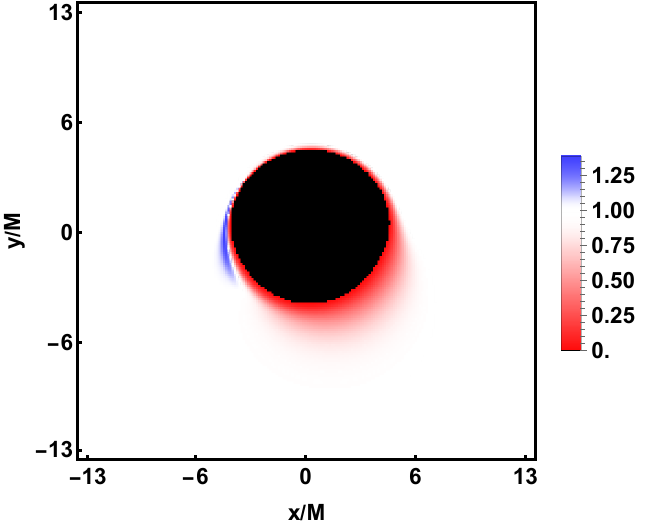}}
\subfigure[\tiny][~$\alpha=0.5,~a=0.3$]{\label{d3}\includegraphics[width=5.4cm,height=4.5cm]{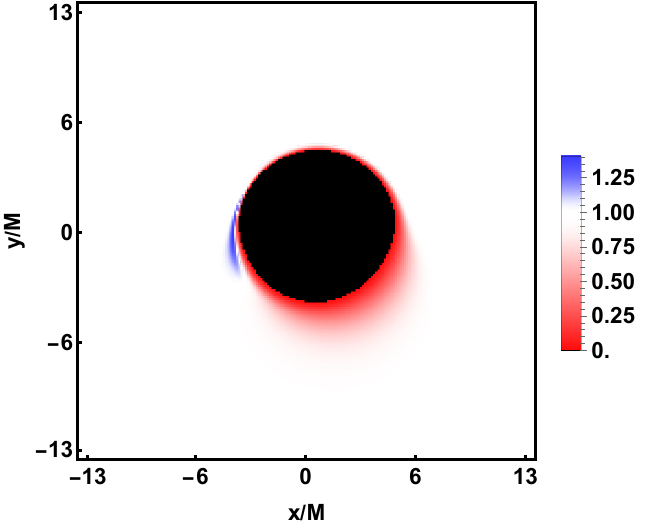}}
\subfigure[\tiny][~$\alpha=0.7,~a=0.001$]{\label{a4}\includegraphics[width=5.4cm,height=4.5cm]{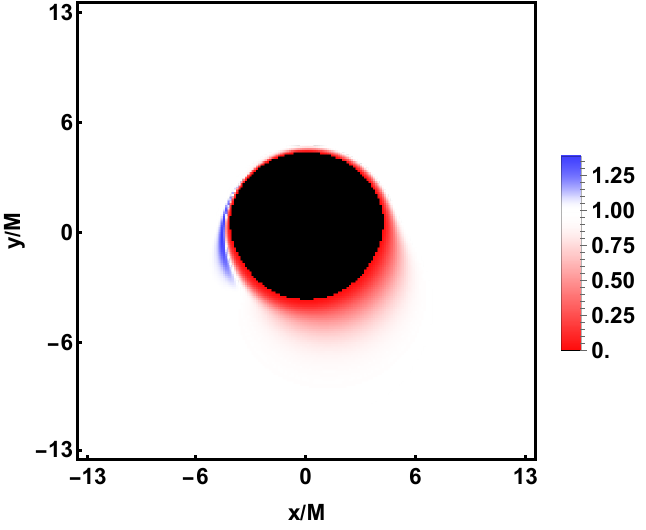}}
\subfigure[\tiny][~$\alpha=0.7,~a=0.1$]{\label{b4}\includegraphics[width=5.4cm,height=4.5cm]{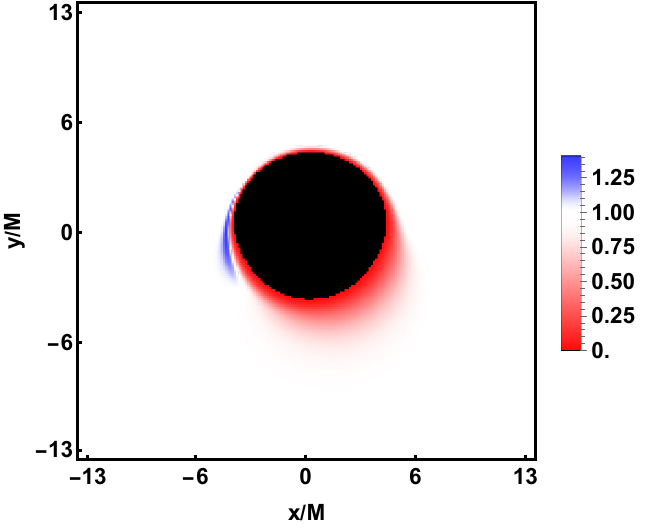}}
\subfigure[\tiny][~$\alpha=0.7,~a=0.3$]{\label{d4}\includegraphics[width=5.4cm,height=4.5cm]{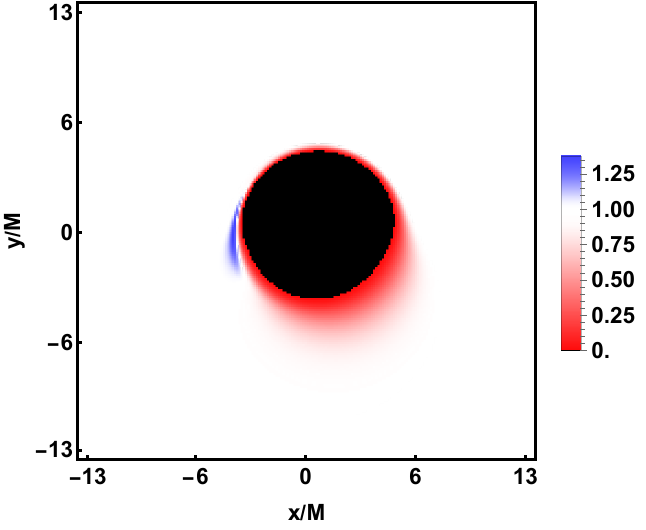}}
\caption{The redshift profiles of lensed images for rotating BHs in $4D$ EGB gravity is examined for different values of $a$ and $\alpha$, while keeping the observer inclination fixed at $\theta_{obs} = 70^\circ$ under prograde flow. The red and blue regions indicate redshifted and blueshifted emission, respectively, whereas the solid black area represents the BH's inner shadow.}\label{prd7}
\end{figure}
Figure~\textbf{\ref{prd7}} shows the optical appearance of the red shift distribution for the lensed images in the prograde flow case, using the same parameter values as defined in Fig.~\textbf{\ref{prd6}}. In all panels, the outer edge of the inner shadow is surrounded by a prominent red crescent-like structure, mainly extending toward the lower-right side of the image. In contrast, the blueshifted region appears as a small petal-like feature on the left side. By fixing the GB parameter $\alpha$ along each row and increasing the spin parameter $a$ from left to right, the blueshifted region becomes less visible, while the redshifted emission becomes more dominant. Similarly, for fixed $a$ along each column, increasing $\alpha$ from top to bottom further suppresses the blueshifted features.
Overall, the lensed images are largely dominated by redshifted emission, with the blue shift contribution remaining comparatively weak.

\begin{figure}
\centering
\subfigure[\tiny][~$\alpha=0.001,~a=0.001$]{\label{a1}\includegraphics[width=5.2cm,height=5cm]{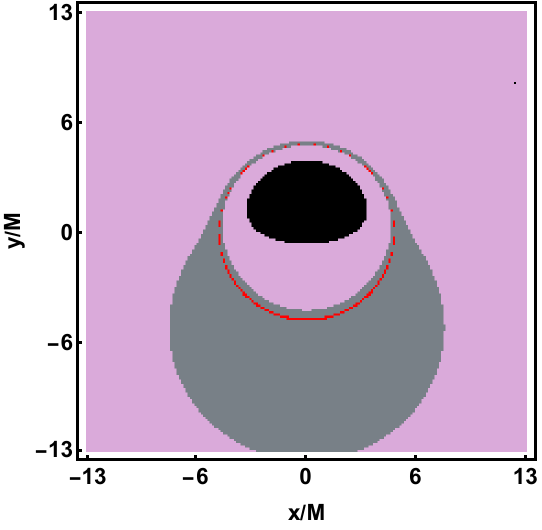}}
\subfigure[\tiny][~$\alpha=0.001,~a=0.1$]{\label{b1}\includegraphics[width=5.2cm,height=5cm]{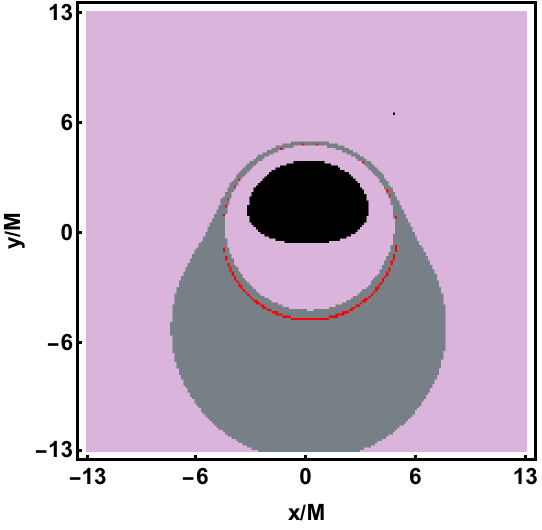}}
\subfigure[\tiny][~$\alpha=0.001,~a=0.3$]{\label{d1}\includegraphics[width=5.2cm,height=5cm]{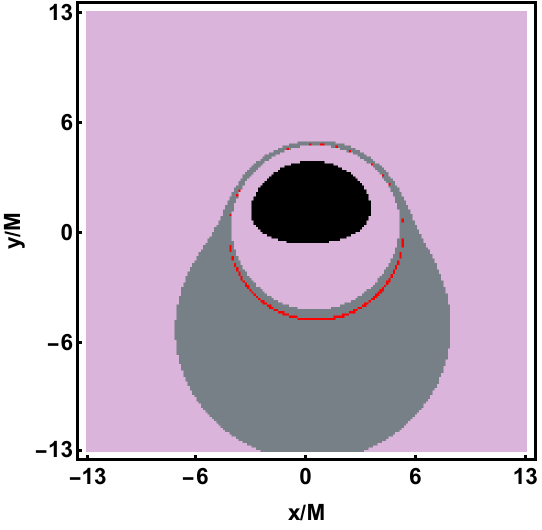}}
\subfigure[\tiny][~$\alpha=0.3,~a=0.001$]{\label{a2}\includegraphics[width=5.2cm,height=5cm]{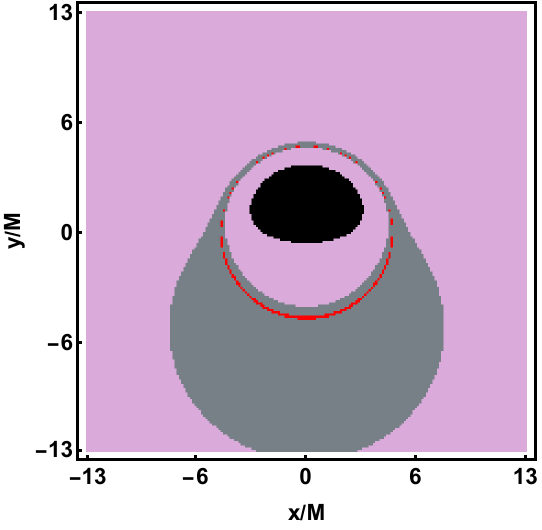}}
\subfigure[\tiny][~$\alpha=0.3,~a=0.1$]{\label{b2}\includegraphics[width=5.2cm,height=5cm]{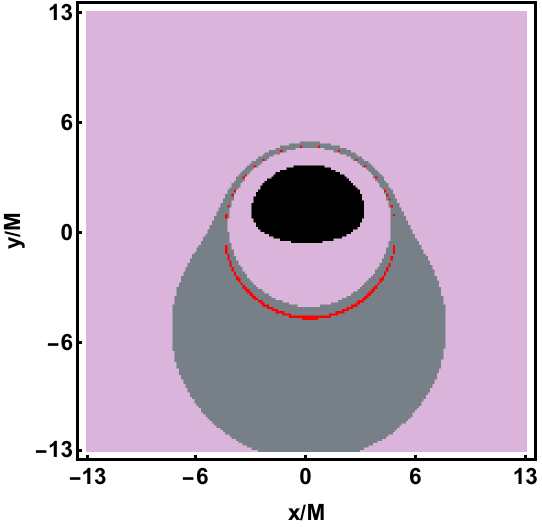}}
\subfigure[\tiny][~$\alpha=0.3,~a=0.3$]{\label{d2}\includegraphics[width=5.2cm,height=5cm]{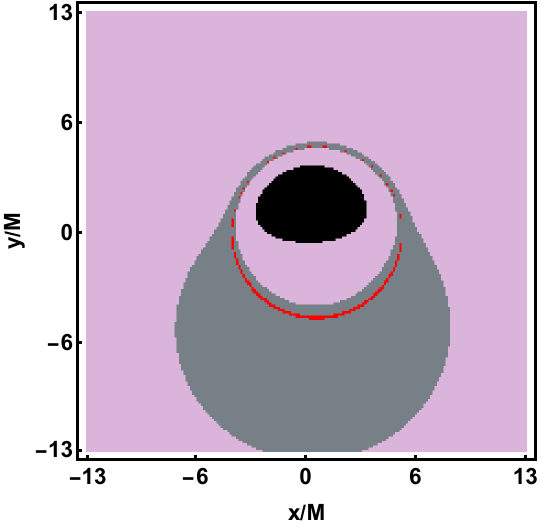}}
\subfigure[\tiny][~$\alpha=0.5,~a=0.001$]{\label{a3}\includegraphics[width=5.2cm,height=5cm]{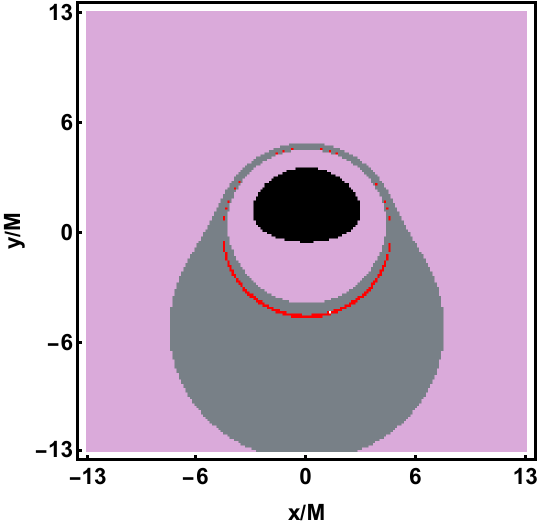}}
\subfigure[\tiny][~$\alpha=0.5,~a=0.1$]{\label{b3}\includegraphics[width=5.2cm,height=5cm]{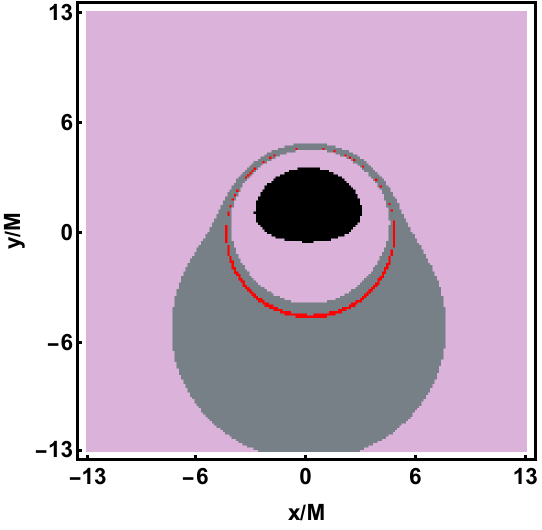}}
\subfigure[\tiny][~$\alpha=0.5,~a=0.3$]{\label{d3}\includegraphics[width=5.2cm,height=5cm]{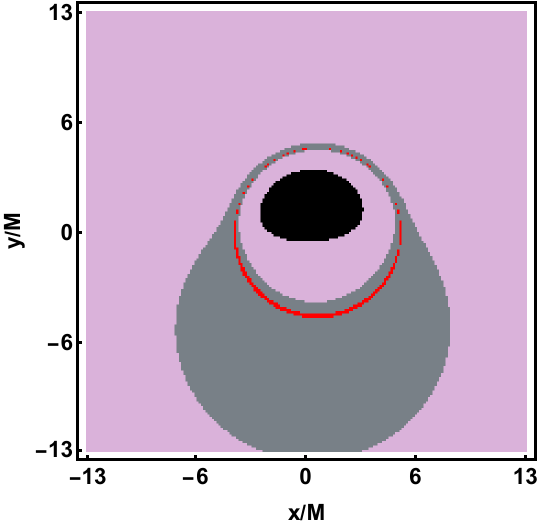}}
\subfigure[\tiny][~$\alpha=0.7,~a=0.001$]{\label{a4}\includegraphics[width=5.2cm,height=5cm]{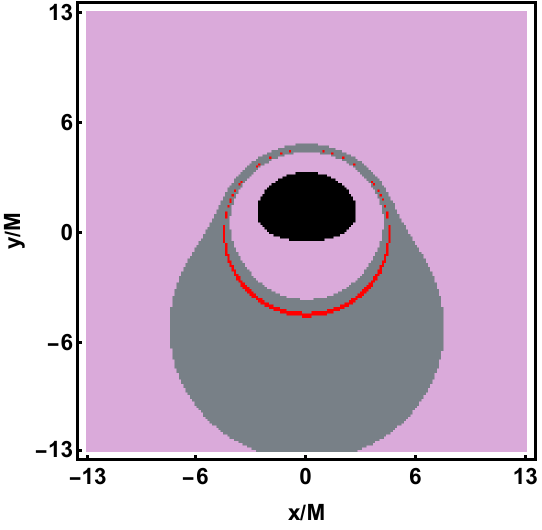}}
\subfigure[\tiny][~$\alpha=0.7,~a=0.1$]{\label{b4}\includegraphics[width=5.2cm,height=5cm]{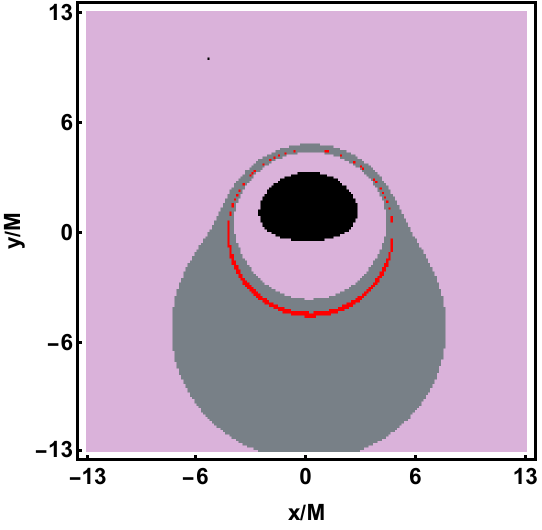}}
\subfigure[\tiny][~$\alpha=0.7,~a=0.3$]{\label{d4}\includegraphics[width=5.2cm,height=5cm]{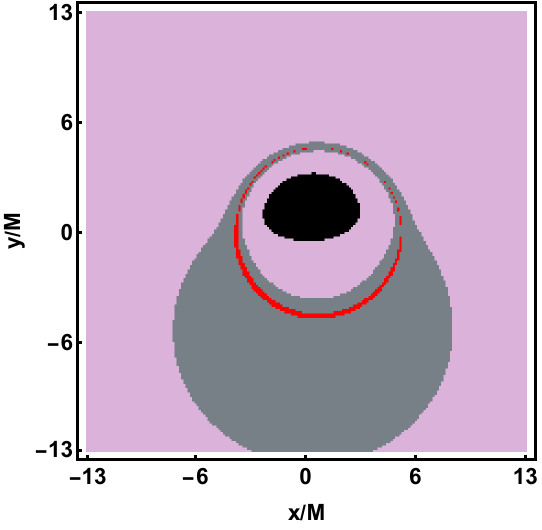}}
\caption{The lensing bands of rotating BHs in $4D$ EGB gravity is examined for different values of $a$ and $\alpha$, while keeping the observer inclination fixed at $\theta_{obs} = 70^\circ$ under prograde flow. The purple, grey, and red regions represent the direct emission, lensed images, and photon ring, respectively, while the solid black area indicates the inner shadow of the BH.}\label{prd8}
\end{figure}
To distinguish between the direct and lensed images of the accretion disk, Fig.~\textbf{\ref{prd8}} presents the corresponding observed flux distributions. The lensing bands are shown in purple, grey, and red, representing the direct emission, lensed images, and photon ring, respectively, while the central black region denotes the inner shadow. In all panels, the lensed bands predominantly appear in the lower half of the image. By fixing the GB parameter $\alpha$ along each row and increasing the spin parameter $a$ from left to right, the lensed structures become more distorted and shift toward the right side of the screen. On the other hand, for fixed $a$ along each column, increasing $\alpha$ from top to bottom causes the lensed bands to contract and slightly move toward the upper region, while the radius of the photon ring gradually decreases. Overall, the photon ring consistently remains enclosed within the purple and grey bands, whereas the inner shadow preserves an almost stable, hat-like profile in all cases. We now examine the influence of the GB parameter $\alpha$ on the visual features of rotating BHs under retrograde thin accretion disk flow. In comparison to the prograde configuration, the effect of gravitational redshift significantly reduces the observed brightness of the shadow. As a result, the overall intensity of the emitted radiation is lowered, making it more difficult to clearly distinguish the lensed images from higher-order images. This reduction in contrast also leads to a slight loss in the sharpness of the photon ring.

\begin{figure}
\centering
\subfigure[\tiny][~$~a=0.001$]{\label{a3}\includegraphics[width=5.4cm,height=4.8cm]{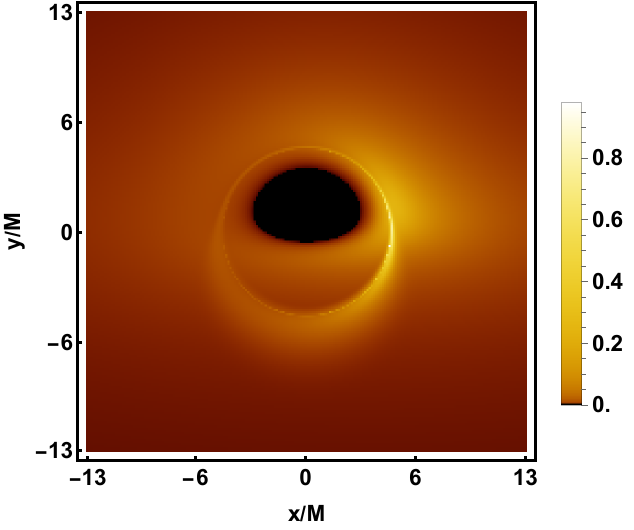}}
\subfigure[\tiny][~$a=0.1$]{\label{b3}\includegraphics[width=5.4cm,height=4.8cm]{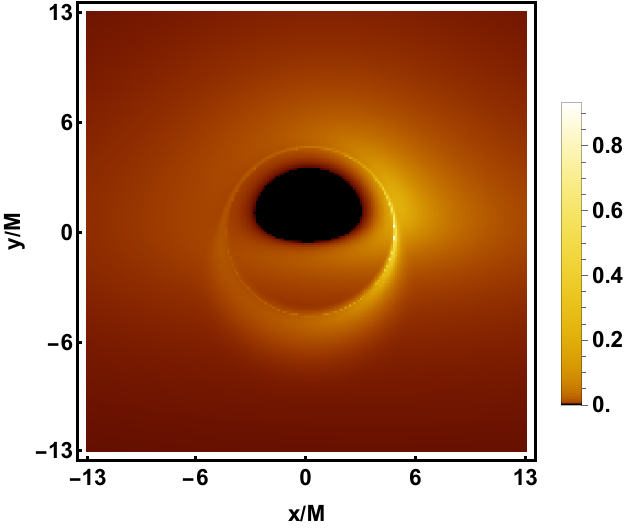}}
\subfigure[\tiny][~$~a=0.3$]{\label{d3}\includegraphics[width=5.4cm,height=4.8cm]{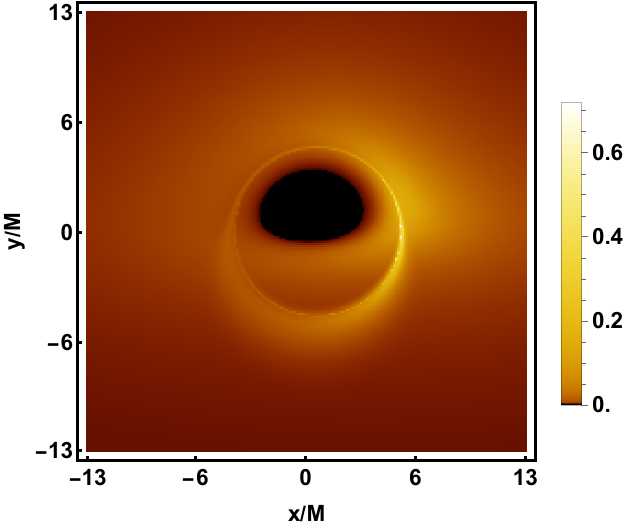}}
\caption{Optical images of rotating BHs in $4D$ EGB gravity is examined for different values of $a$, with fixed $\alpha = 0.5$ and $\theta_{\text{obs}} = 70^\circ$ under retrograde flow. The central dark region represents the BH event horizon, while the surrounding bright circular ring indicates the location of the photon ring.}\label{prd9}
\end{figure}
From Fig.~\textbf{\ref{prd9}}, it can be seen that a bright emission region is present in the lower right portion of the image. As the spin parameter $a$ increases, this luminous structure gradually contracts and shifts slightly toward the upper region of the screen. 
Moreover, a crescent-like bright feature appears in the upper right part of the image, whose intensity becomes more pronounced with increasing $a$. This behavior can be attributed to the fact that the material located near the event horizon contributes less to the observed brightness, whereas the emission from matter confined to the equatorial plane enhances the overall optical depth during the imaging process.

\begin{figure}
\subfigure[\tiny][~$~a=0.001$]{\label{a3}\includegraphics[width=5.4cm,height=4.5cm]{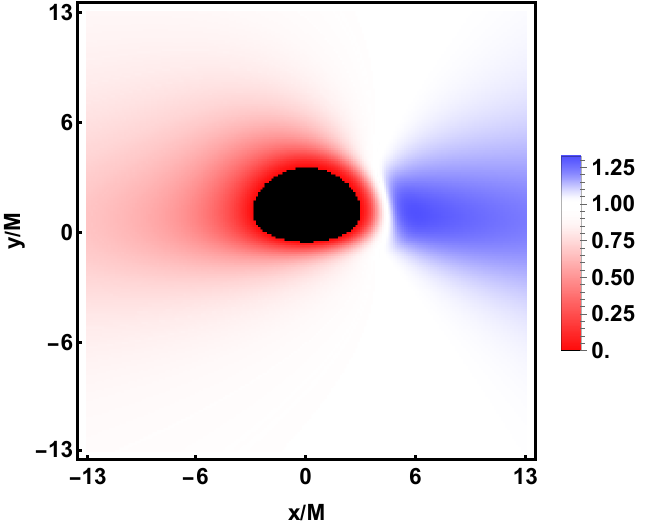}}
\subfigure[\tiny][~$a=0.1$]{\label{b3}\includegraphics[width=5.4cm,height=4.5cm]{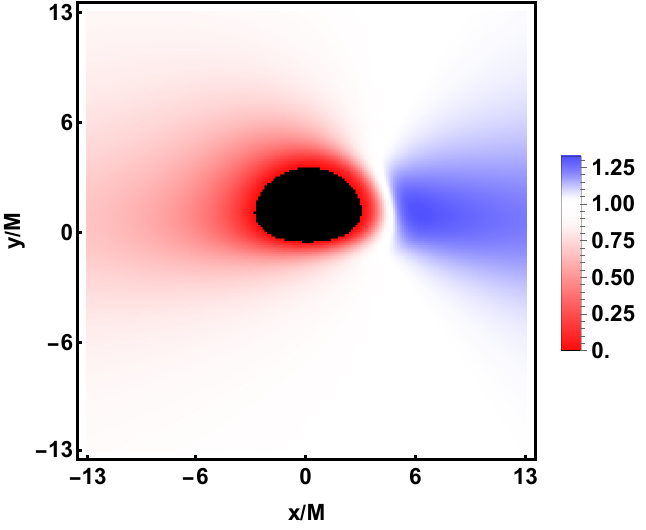}}
\subfigure[\tiny][~$~a=0.3$]{\label{d3}\includegraphics[width=5.4cm,height=4.5cm]{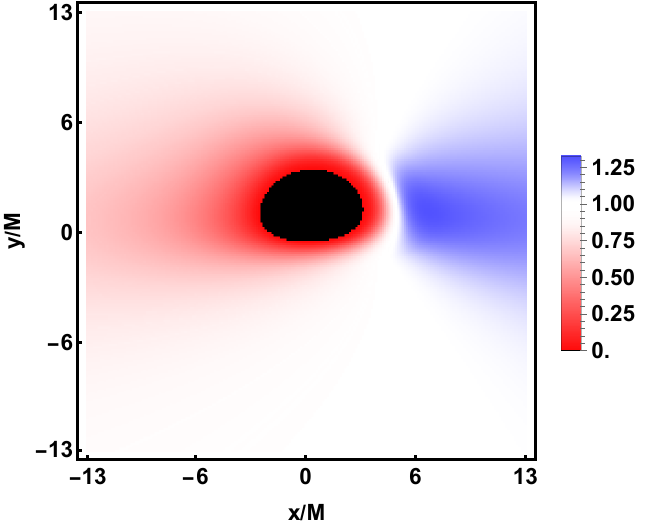}}
\caption{The red-shift distribution for the direct images of rotating BHs in $4D$ EGB gravity is examined for different values of $a$, with fixed $\alpha=0.5$ and $\theta_{\text{obs}} = 70^\circ$ under retrograde flow. The red and blue regions correspond to redshifted and blueshifted emission, respectively, while the central black area represents the inner shadow of the BH.}\label{prd10}
\end{figure}
Figure~\textbf{\ref{prd10}} illustrates the red shift distribution for direct images under retrograde flow for different values of the spin parameter $a$. In contrast to the prograde case, the orientation of the red and blueshifted regions is reversed. The redshifted emission extends over a larger area and more clearly surrounds the inner shadow of the BH. However, variations in $a$ have only a minor effect on the overall red shift distribution, indicating a relatively weak dependence on this parameter in the direct imaging case.
\begin{figure}
\centering
\subfigure[\tiny][~$~a=0.001$]{\label{a3}\includegraphics[width=5.4cm,height=4.5cm]{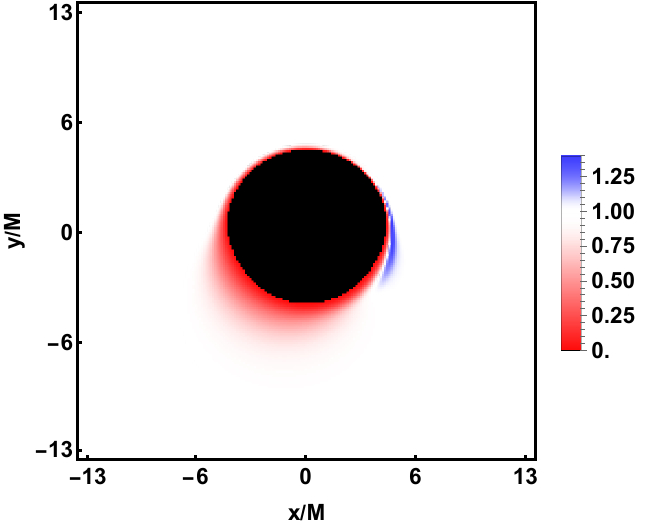}}
\subfigure[\tiny][~$a=0.1$]{\label{b3}\includegraphics[width=5.4cm,height=4.5cm]{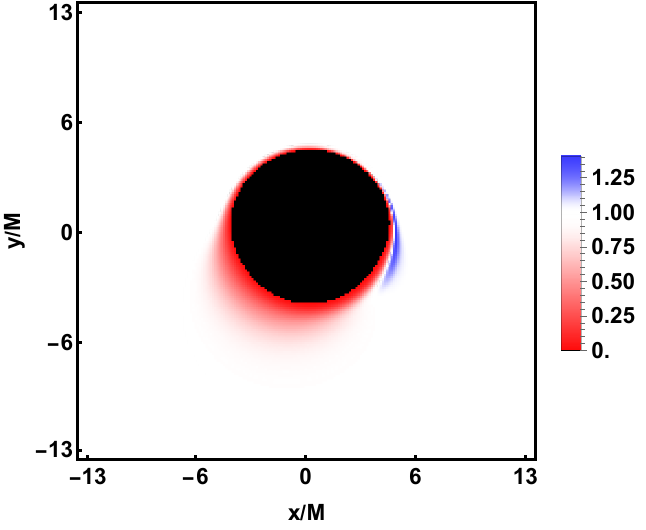}}
\subfigure[\tiny][~$~a=0.3$]{\label{d3}\includegraphics[width=5.4cm,height=4.5cm]{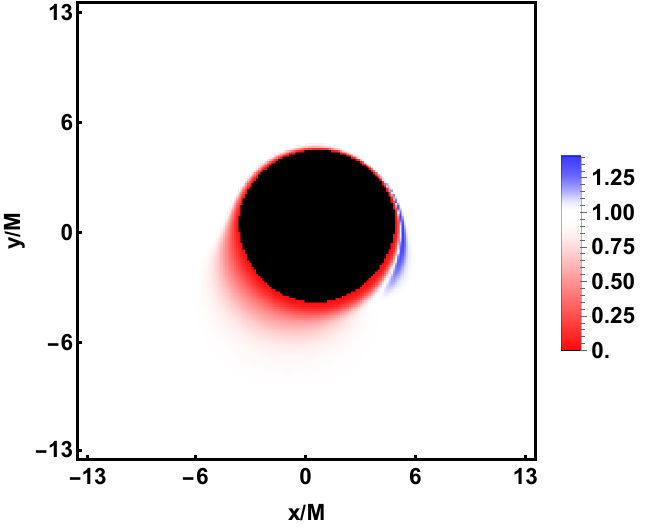}}
\caption{The red-shift distribution for the lensed images of rotating BHs in $4D$ EGB gravity is examined for different values of $a$, with fixed $\alpha=0.5$ and $\theta_{\text{obs}} = 70^\circ$ under retrograde flow. The red and blue regions correspond to redshifted and blueshifted emission, respectively, while the central black area represents the inner shadow of the BH.}\label{prd11}
\end{figure}
Figure~\textbf{\ref{prd11}} presents the corresponding red shift distribution for the lensed images. In this case, the redshifted region is mainly confined to the lower-left portion of the screen, and its extent gradually decreases as $a$ increases. Meanwhile, a thin blueshifted strip appears near the inner shadow on the right side, remaining nearly unchanged with varying $a$. This suggests that the influence of the spin parameter is more noticeable in modifying the extent of redshifted regions in lensed images than in direct ones.

\begin{figure}
\centering
\subfigure[\tiny][~$~a=0.001$]{\label{a3}\includegraphics[width=5.2cm,height=5cm]{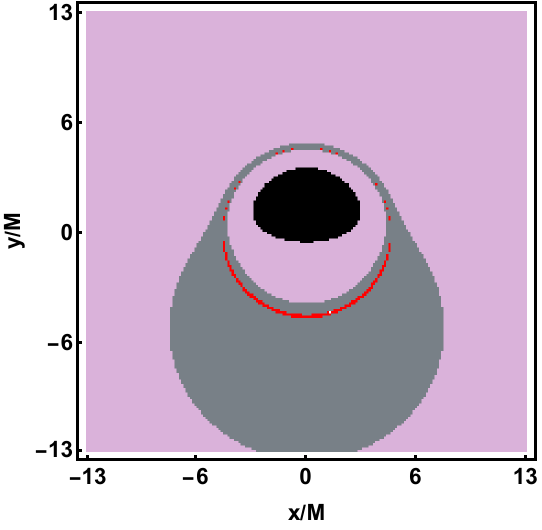}}
\subfigure[\tiny][~$a=0.1$]{\label{b3}\includegraphics[width=5.2cm,height=5cm]{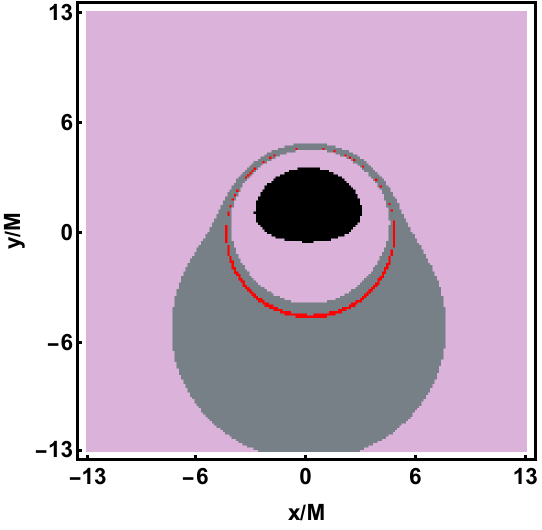}}
\subfigure[\tiny][~$~a=0.3$]{\label{d3}\includegraphics[width=5.2cm,height=5cm]{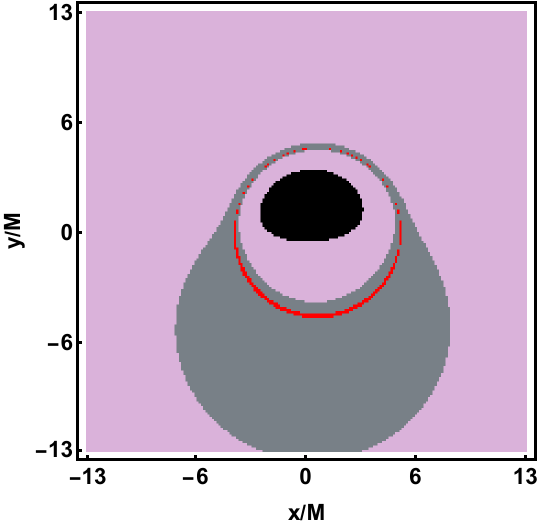}}
\caption{The lensing band of rotating BHs in $4D$ EGB gravity is examined for different values of $a$, with fixed $\alpha = 0.5$ and $\theta_{\text{obs}} = 70^\circ$ under retrograde flow. The purple, grey, and red regions represent the direct emission, lensed images, and photon ring, respectively, while the central black area corresponds to the inner shadow of the BH.}\label{prd12}
\end{figure}

Figure~\textbf{\ref{prd12}} presents the optical appearance of the lensing bands for the BH model considered under retrograde accretion flow, with varying values of $a$. It is observed that as $a$ increases, the lensed structures progressively contract and shift toward the lower-right region of the image.
Meanwhile, the inner shadow maintains a nearly stable hat-like shape with only minor changes, and the photon ring consistently remains enclosed within the direct and lensed emission bands across all cases.

\subsection{Observational Bounds with EHT Data}
\begin{figure}
\centering
\subfigure[\tiny][~$a=0.1$]{\label{a1}\includegraphics[width=8cm,height=5.5cm]{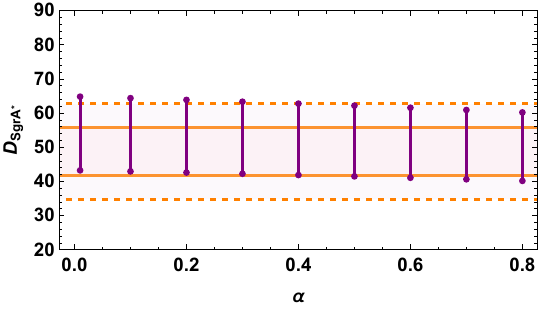}}
\subfigure[\tiny][~$a=0.1$]{\label{b1}\includegraphics[width=8cm,height=5.5cm]{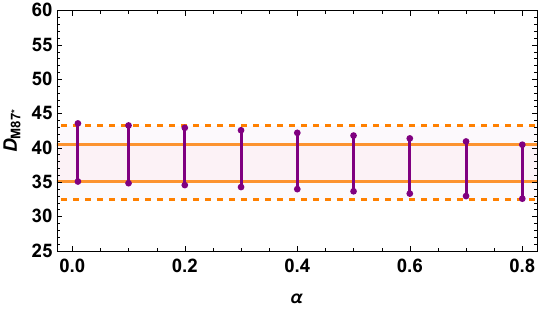}}
\caption{The profiles represent the variation of the approximated shadow angular diameter of Sgr A$^{\ast}$ and ${M87^{\ast}}$ in left and right panels, respectively. The solid and dashed orange lines denote the $1\sigma$ and $2\sigma$ confidence levels, respectively, while the purple bands represent the estimated range. In both cases, the rotation parameter is fixed at $a=0.1$.}
\label{est}
\end{figure}

Now, we constrain the GB coupling parameter $\alpha$ employing the observational data provided by the EHT for M87$^{\ast}$ and Sgr A$^{\ast}$. To achieve this, we compute the angular diameter of the BH shadow and perform a comparison with the corresponding observed values for these sources. The angular diameter $D$ of the shadow is defined in terms of the shadow radius as
$D = 2 \mathcal{R}_{d} \frac{\mathcal{M}}{\mathcal{D}_{obs}}$,
where $\mathcal{R}_{d}$ denotes the dimensionless shadow radius evaluated at the observer's location, $\mathcal{M}$ is the BH mass, and $\mathcal{D}_{\mathrm{obs}}$ represents the distance between the observer and the BH. In the present analysis, $\mathcal{R}_{d}$ is obtained from the shadow cast by a rotating BH in $4$D EGB gravity surrounded by a thin accretion disk, and it explicitly depends on the GB parameter $\alpha$ along with the variations of $a$. Therefore, by comparing the theoretically estimated angular diameter with the EHT measurements, we are able to place bounds on the allowed values of $\alpha$ \cite{62, 64}. Mathematically, the angular diameter can be written as
\begin{equation}\label{24}
D=2\times9.87098\mathcal{R}_{d}\big(\frac{\mathcal{M}}{M_{\odot}}\big)\big(\frac{1\text{kpc}}{\mathcal{D}_{obs}}\big)\mu
\end{equation}
Using Eq.~(\ref{24}), the theoretical values of the angular diameter can be evaluated for different choices of the model parameters and subsequently compared with the available observational limits. For M87$^{\ast}$, the source is located at a distance of $\mathcal{D}_{obs} = 16.8~\mathrm{kpc}$, and its mass is estimated to be $\mathcal{M} = (6.5 \pm 0.7)\times 10^{6} M_{\odot}$. The corresponding observed angular diameter of the shadow is reported as $D_{M87^{\ast}} = (37.8 \pm 2.7)\,\mu\mathrm{as}$ \cite{65}. Similarly, for Sgr A$^{\ast}$, the distance from Earth is approximately $\mathcal{D}_{\mathrm{obs}} = 8~{kpc}$, while its mass is constrained to be $\mathcal{M} = (4.0^{+1.1}_{-0.6})\times 10^{6} M_{\odot}$. The measured angular diameter of its shadow is given by $D_{\mathrm{Sgr~A}^{\ast}} = (48.7 \pm 7)\,\mu\mathrm{as}$ \cite{66}. In this analysis, we illustrate the behavior of the shadow angular diameter $D$ as a function of the GB coupling parameter $\alpha$ for M87$^{\ast}$ and Sgr A$^{\ast}$, shown in the left and right panels of Fig. \textbf{\ref{est}}, respectively, while keeping the spin parameter fixed at $a = 0.1$. From these panels, it can be observed that the theoretical predictions lie predominantly within the observational confidence intervals $1\sigma$ and $2\sigma$. This agreement suggests that the values of $\alpha$ considered in the model are consistent with the constraints inferred from both M87$^{\ast}$ and Sgr A$^{\ast}$ observations. Therefore, current astronomical data provide meaningful bounds on the GB parameter for these supermassive BHs. Furthermore, the rotation parameter $a$ mainly influences the distortion or deviation of the shadow from circularity, while its effect on the overall size of the shadow remains comparatively small. For this reason, its role is not emphasized in the present discussion. Future high-precision observations of BH shadows are expected to further tighten the constraints on the GB coupling parameter $\alpha$.
\section{Discussion and Conclusion}
In recent years, the investigation of BH spacetime through shadow imaging and accretion-disk models has gained significant attention, especially with the rise of high-resolution observations. Within the framework of $4$D EGB gravity, our analysis highlights how the GB parameter $\alpha$ modifies the observable features ofBH images. In particular, the presence of a photon ring surrounding a central dark region provides key information about the underlying geometry and the behavior of radiation near the event horizon. Employing thin accretion disk models, we demonstrate that variations in $\alpha$ and spin parameter $a$ lead to noticeable changes in the shadow size, distortion, and brightness distribution. Our results show that, from the analysis of the BH horizon function $\Delta(r)$, the horizon radii are influenced by both $\alpha$ and $a$. Specifically, the horizon radius decreases with increasing $a$, while the separation between the curves slightly grows as $\alpha$ increases. Moreover, as both $\alpha$ and $a$ increase, the curves shift upward, and the intersection with the radial axis disappears, indicating the absence of two distinct event horizons in this regime. This implies that the event horizon vanishes in this regime, and the solution no longer corresponds to a BH. The shadow contours indicate that for small values of $a$, as shown in the left panel of Fig.~\textbf{\ref{prd2}}, the shadow remains nearly symmetric with only slight deviations from a perfect circular shape. In contrast, the right panel of Fig.~\textbf{\ref{prd2}} demonstrates that for larger values of $a$, the contours become noticeably distorted. In particular, as $\alpha$ decreases, the shadow shifts toward the left side of the screen, while a clear flattening appears along the right side.

Based on the celestial light source model, we analyze the optical appearance of BH shadows for different values of the rotation parameter $a$ and the GB parameter $\alpha$. The results show that both parameters play distinct roles in shaping the shadow. For very small values of $a$ and $\alpha$, the shadow remains nearly circular and symmetric. As the spin parameter $a$ increases, the shadow becomes distorted and shifts due to enhanced frame-dragging effects, leading to a noticeable deformation in its shape. On the other hand, increasing $\alpha$ mainly affects the size and slightly contracts the shadow radius while preserving its overall geometry. In the thin accretion disk scenario, the prograde flow reveals that variations in $\alpha$ produce a smooth, hat-like inner shadow with minor deformation, whereas increasing $a$ significantly reduces the shadow size and generates a bright crescent-like region due to relativistic Doppler effects. Overall, the combined influence of $a$ and $\alpha$ can be clearly distinguished through their impact on the shadow shape, size, and brightness distribution in $4$D EGB gravity. We further examine the red shift distribution of accreting matter for both direct and lensed images. In direct images, the blueshifted region appears on the left side, while the redshifted region dominates the right side of the screen. It is observed that increasing the spin parameter $a$ significantly alters the red shift distribution, reducing its intensity, whereas $\alpha$ introduces a mild contraction in the red shift region. 

In all cases, the blueshifted component remains comparatively weaker and occupies a smaller area. For lensed images, the boundary of the inner shadow is surrounded by a distinct red crescent-like feature extending toward the lower-right region, while the blueshift appears as a small petal on the left side and remains suppressed. Additionally, the lensing bands show that increasing $a$ enhances their deformation and shifts them toward the lower-right side, whereas increasing $\alpha$ causes a gradual contraction and slight upward displacement. Overall, the spin parameter mainly governs asymmetry and intensity, while $\alpha$ primarily affects the size and extent of the observed features. In the retrograde accretion scenario, a bright emission region is observed in the lower-right part of the image, which gradually contracts and shifts slightly upward with increasing $\alpha$. A crescent-like bright feature also appears in the upper-right region, becoming more pronounced as $\alpha$ increases. The redshift distribution for both direct and lensed images shows that the redshifted emission extends over a larger area and more clearly surrounds the inner shadow, while its dependence on $\alpha$ remains weak. For lensed images, the redshifted region is mainly confined to the lower-left side, whereas a narrow blueshifted strip appears near the inner shadow on the right and remains nearly unchanged. Furthermore, the lensing bands shrink toward the lower-right region with increasing $\alpha$, while the photon ring consistently stays enclosed between the direct and lensed components. These results indicate that observational features can be used to place constraints on the GB parameter $\alpha$ in $4$D EGB gravity.

Overall, the present study demonstrates that the main physical parameters of a BH have significant effects on its shadow features in the celestial light sphere and geometrically thin accretion flow models. Such models provide a more realistic description of extreme astrophysical environments around rotating BHs. By combining shadow images, intensity distributions and redshift configurations, one can obtain a more comprehensive
characterization of the radiation properties and spacetime structure around BHs through effective implementations. In future studies, it would be worthwhile to compare the imaging signatures of BHs with those of other
compact objects, such as neutron stars and boson stars, in order to explore observable differences among different gravitational frameworks. These efforts may provide effective theoretical background for forthcoming high-resolution astronomical observations.

\vspace{.25cm}
\end{document}